\documentclass[a4paper]{nature}

\usepackage{color}
\usepackage{url}
\usepackage{verbatim}
\usepackage{multirow}
\usepackage{xspace}
\usepackage{graphicx}
\usepackage{amsmath} 
\usepackage{bbm}
\usepackage{amssymb}
\usepackage{booktabs} 
\usepackage{times} 
\usepackage{lscape}
\usepackage[left=1in,right=1in,bottom=1.1in,top=1in]{geometry}
\usepackage{enumitem}
\usepackage[table,xcdraw]{xcolor}
\usepackage[normalem]{ulem}
\usepackage{xcolor}
\usepackage[innercaption]{sidecap}
\usepackage{lineno}

\usepackage{sectsty}   % Allows font changes to section headers
\allsectionsfont{\sffamily}  % Apply Helvetica (sans-serif) to all section headers

\usepackage[colorlinks,citecolor=blue,urlcolor=magenta,filecolor=red]{hyperref}
\usepackage[margin=-1cm, labelfont=bf, font=footnotesize]{caption}
\usepackage{setspace}
\usepackage[ruled,vlined]{algorithm2e}
\SetKwRepeat{Do}{do}{while}

\newcommand{\revision}[1]{{#1}}

\usepackage[font=footnotesize, labelfont={sf,bf}]{caption}
\newcommand{\textsbf}[1]{\textsf{\textbf{#1}}}
\newcommand{\xhdrt}[1]{\vspace{5mm}\noindent{{\textsbf{#1}}}\vspace{2.5mm}}
\newcommand{\xhdr}[1]{\vspace{5mm}\noindent{{\textsbf{#1.}}}}

\newcommand{\hide}[1]{}

\let\oldnl\nl
\newcommand{\nonl}{\renewcommand{\nl}{\let\nl\oldnl}}  
\usepackage[normalem]{ulem}
\usepackage{comment}
\usepackage{bm}
\usepackage{bbm}
\graphicspath{{./FIG/}}

\newcommand{\todo}[1]{{\color{red} [TODO]}}

\newcommand{\name}{\textsc{Madrigal}\xspace}

\title{\begin{center}
Multimodal AI predicts clinical outcomes of drug combinations from preclinical data
\vspace{-10mm}
\end{center}}

\author  
{\begin{center}{\small   
Yepeng Huang$^{1,2}$, Xiaorui Su$^{1}$, Varun Ullanat$^{1}$, Intae Moon$^{1}$, Ivy Liang$^{3}$, \\ Lindsay Clegg$^{4}$, Damilola Olabode$^{5}$, Ruthie Johnson$^{1}$, Nicholas Ho$^{6}$, Megan Gibbs$^{5}$, \\ Alexander Gusev$^{8,10}$, Bino John$^{7}$, and Marinka Zitnik$^{1,9,10,11,\ddag}$} \\[2mm]
\footnotesize{$^{1}$Department of Biomedical Informatics, Harvard Medical School, Boston, MA, USA} \\  
\footnotesize{$^{2}$Program in Biological and Biomedical Sciences, Harvard Medical School, Boston, MA, USA} \\  
\footnotesize{$^{3}$Harvard College, Cambridge, MA, USA} \\
\footnotesize{$^{4}$Clinical Pharmacology and Quantitative Pharmacology, Clinical Pharmacology \& Safety Sciences, R\&D, AstraZeneca, Gaithersburg, MD, USA}\\
\footnotesize{$^{5}$Clinical Pharmacology and Quantitative Pharmacology, Clinical Pharmacology \& Safety Sciences, R\&D, AstraZeneca, Waltham, MA, USA} \\
\footnotesize{$^{6}$Program in Computational Biology, Carnegie Mellon University, Pittsburgh, PA, USA}\\
\footnotesize{\!\!\!\!\!\!\!\!\!\!$^{7}$Imaging and Data Analytics, Clinical Pharmacology \& Safety Sciences, R\&D, AstraZeneca, Waltham, MA, USA \!\!\!\!\!\!\!\!\!\!}\\
\footnotesize{\!\!\!\!\!\!\!\!\!\!$^{8}$Department of Medical Oncology, Dana-Farber Cancer Institute and Harvard Medical School, Boston, MA, USA \!\!\!\!\!\!\!\!\!\!} \\
\footnotesize{$^{9}$Kempner Institute for the Study of Natural and Artificial Intelligence, Harvard University, Allston, MA, USA \!\!\!\!} \\
\footnotesize{$^{10}$Broad Institute of MIT and Harvard, Cambridge, MA, USA} \\
\footnotesize{$^{11}$Harvard Data Science Initiative, Cambridge, MA, USA}\\
\footnotesize{$^{\ddag}$Corresponding author: marinka@hms.harvard.edu}
\end{center}
}

\usepackage{pdfpages}
\begin{document}

\maketitle

% \linenumbers

{\customspacing{1}
\begin{abstract}
\section*{Abstract}
Predicting clinical outcomes from preclinical data is essential for selecting safe and effective drug combinations and for reducing late-stage failures. AI models use molecular structure and target annotations, and do not leverage the perturbation readouts that report how a compound acts in a cellular context.
Here we introduce \name, a multimodal AI model that learns from structural, pathway, cell-viability, and transcriptomic data. \name aligns these modalities across 21,842 compounds into a shared latent space and predicts combination outcomes even for drugs observed in only a subset of the data modalities.
Trained on 158 expert-curated and 795 patient-reported combination outcomes, \name outperforms single-modality and state-of-the-art multimodal methods. Ablations show that modality alignment and multimodal input each improve predictive performance.
\name predicts elevated risk for combinations that share membrane transporters. In head-to-head trials that compare two combination arms, the arm with the higher observed incidence of neutropenia, anemia, alopecia, or hypoglycemia receives the higher predicted risk in 25 of 28 comparisons. In MASH, \name ranks resmetirom among the candidates with favorable predicted safety when paired with type 2 diabetes drugs.
\revision{\name also improves adverse-event prediction in a longitudinal patient cohort and an independent oncology cohort} and predicts efficacy in primary acute myeloid leukemia samples and patient-derived xenografts.

\end{abstract}
}

\clearpage

\customspacing{1.2}

\section*{Main}

Combination therapies are standard in modern treatment, where complementary mechanisms and dose reduction allow a regimen to improve efficacy and to lower the toxicity of an individual drug~\cite{Palmer_2017_combo,Jin2023rationalcancercombo,Ratziu_2023_rational_combo_nash}. The same co-administration can increase the risk of adverse drug reactions driven by drug-drug interactions (DDIs), and benefit and risk frequently arise in the same regimen. Infliximab plus azathioprine improves corticosteroid-free remission in Crohn's disease compared with monotherapy but is associated with infections and other adverse events~\cite{colombel2010infliximab}, and nivolumab with chemotherapy prolongs overall survival in advanced esophageal squamous cell carcinoma at the cost of higher treatment-related toxicities~\cite{doki2022nivolumab}. These risks are magnified in vulnerable populations, including cancer survivors~\cite{Murphy2018} and patients with chronic~\cite{Menditto2019} and neurological diseases~\cite{Csoti2019}. 

Developing combination therapies that are both effective and safe is difficult due to the combinatorial explosion of drug combinations and heterogeneous clinical effects across patients and indications~\cite{Sun2015,sun2016drug}. The challenge is most acute for novel compounds in preclinical or early clinical development, where critical information that emerges only later, including comprehensive safety profiles, long-term efficacy, and pharmacokinetics, is missing. This data sparsity limits prediction of drug interactions and therapeutic responses from preclinical data alone. Models that predict combination outcomes from multimodal preclinical data are therefore essential to improve clinical success, minimize patient risk, and avoid unnecessary studies~\cite{reality_check_2023}.

Multiple preclinical modalities offer complementary views of combination effects: molecular structures, mechanisms of action, and cell-based perturbation readouts. While structure is universally available and predictive of bioactivity~\cite{huang2021therapeutics, swanson2024admet}, it often cannot capture combination behavior driven by context-dependent pharmacodynamics beyond intrinsic molecular structure~\cite{karunajeewa2008trial} (Supplementary Fig.~\ref{fig:supp_fig_1}{b}), motivating broader representations of drug action~\cite{jaaks2022effective}. \revision{Target and pathway annotations encode mechanism of action directly, with molecular interaction networks predicting polypharmacy side effects~\cite{Zitnik2018} and shared metabolic enzymes or transporters underlying established pharmacokinetic interactions~\cite{Shi_2024_transportome}, yet such annotations are curated and thus incomplete for novel compounds and blind to off-target activity.}
High-throughput perturbation data, including transcriptional responses in cell lines~\cite{cmap} and cell-viability profiles~\cite{prism} following chemical perturbations, encode complex biological consequences of drugs~\cite{Lamb2006, Cohen2008, Molinelli2013} but remain underutilized in predictive modeling despite their importance for synergy and safety~\cite{Lukačišin2019, Wu2023}. Transcriptional phenotypes report drug-protein interactions and pathway-level shifts in cellular activity~\cite{Iorio2010, Jang2021, Pham2021, Pham2022}, whereas cell-viability data reveal pathway dependencies across tissues and cell types and help uncover gene functions relevant to drug action~\cite{Barretina2012, Rees2016, Pan2022}. These data modalities are critical for target prioritization~\cite{Oberlick2019}, detecting adverse drug interactions~\cite{Corsello2020, Raghavan2021, Hahn2021}, and managing polypharmacy in patients with comorbidities.

A central barrier is the ``missing-modality'' problem~\cite{Wu_Jastrze_2022, Huang2022,ektefaie2023multimodal} (Fig.~\ref{fig:figure_1}c, Supplementary Fig.~\ref{fig:supp_fig_1}{c}): for novel or early-stage compounds, and even some approved drugs, critical modalities (e.g., pathway annotations, perturbational readouts, toxicity evidence) are absent. Models that assume complete data during training either exclude such compounds or impute them in ways that distort modality structure, degrading generalization to real-world settings. The impact is most severe for novel compounds lacking pathway labels and for early experimental drugs with sparse safety signals. Thus, models must learn from heterogeneous, partially observed modality sets and remain robust to missing inputs at both training and inference, leveraging whatever evidence is available without requiring complete profiles.

\revision{Here, we introduce \name, a multimodal AI model that predicts safety outcomes of drug combinations (Fig.~\ref{fig:figure_1}a).} A separate encoder maps each modality of a drug to a representation, and contrastive learning aligns these representations to the structure representation in a shared latent space, so that the modalities observed for a drug inform the modalities the drug lacks (Fig.~\ref{fig:figure_1}b,c). An attention bottleneck fuses the available representations into a single drug embedding, and a decoder scores a pair of embeddings against each safety outcome (Supplementary Fig.~\ref{fig:supp_fig_1}a). 
\name predicts higher risk of serum-level and excretion-related outcomes for combinations whose drugs share a membrane transporter, and it ranks pairs such as doxycycline with tacrolimus among the most concerning. In head-to-head trials that compare two combination arms, the arm with the higher observed incidence of alopecia, anemia, hypoglycemia, or neutropenia receives the higher predicted risk in 25 of 28 comparisons. In metabolic dysfunction-associated steatohepatitis (MASH), \name ranks resmetirom among the candidates with favorable predicted safety when paired with type 2 diabetes (T2D) drugs. \name also predicts effective combinations in primary acute myeloid leukemia~\cite{Bottomly_2022_BeatAML,Eide_2023_Venetoclax} and in patient-derived xenografts~\cite{Gao_2015_pdx}, and it predicts patient adverse-event risk in a longitudinal event-time cohort~\cite{ehrshot} and in an independent oncology cohort.

\section*{Results}
\xhdrt{\name multimodal AI model for predicting drug combination outcomes}

\begin{figure}[t]
% \centering
% \vspace*{-2cm}
\makebox[\linewidth]{
        \includegraphics[width=\linewidth]{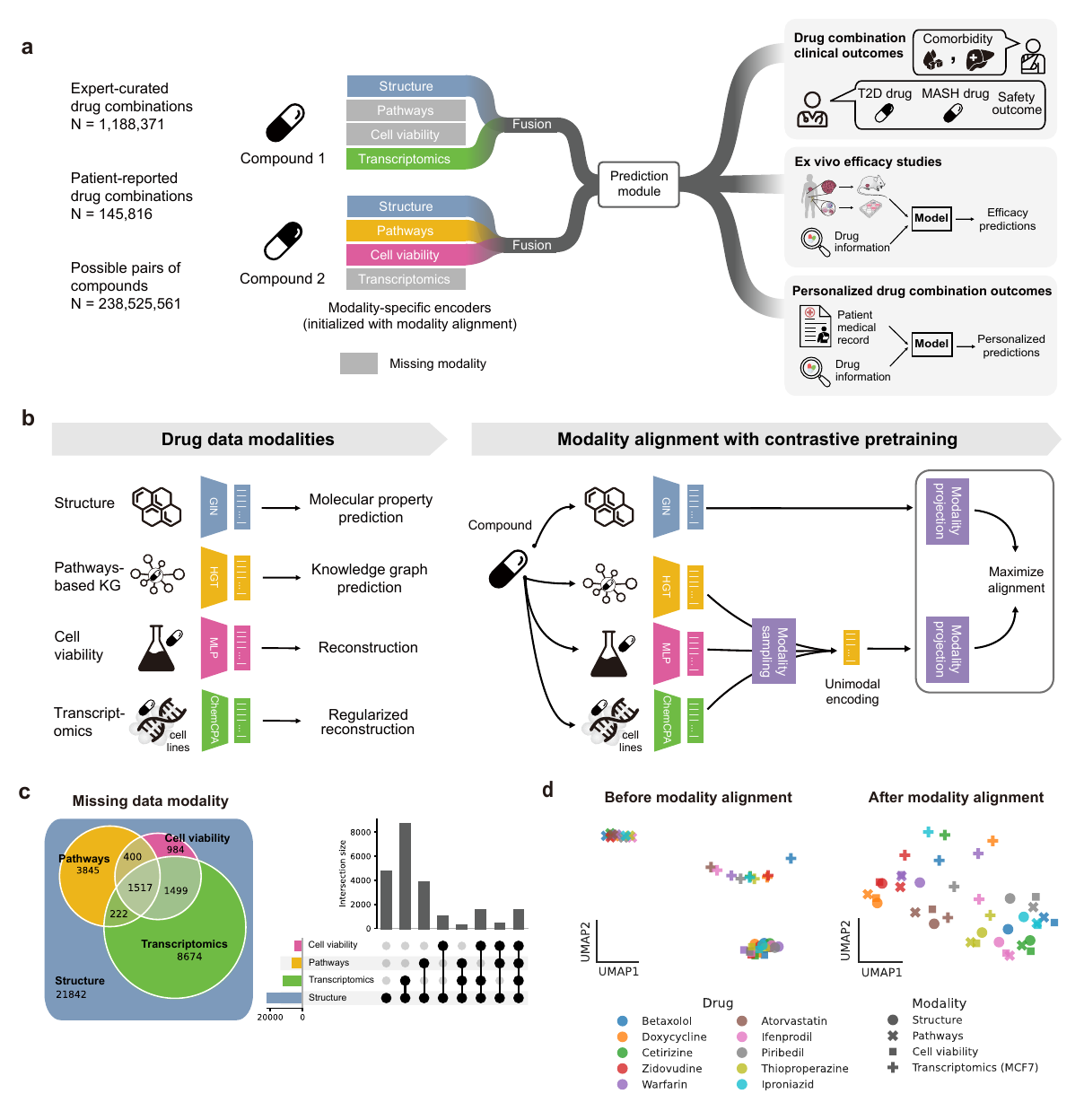}
    }
\caption{\textsbf{\name integrates multimodal preclinical data to predict clinical outcomes of drug combinations.} 
\textsbf{a,} Overview of \name. Comprising modality-specific encoders, a fusion module, and a prediction module, \name is trained on expert-curated and patient-reported drug combination datasets to predict clinical outcomes. Modalities are fused with attention bottleneck modules (Methods and Supplementary Note~\ref{note:architecture}). The model enables three key applications: 
prediction of polypharmacy safety outcomes, 
efficacy prediction in patient-derived models,
and personalized drug combination outcome predictions in patients.
\textsbf{b,} Overview of data modalities and the modality alignment framework. 
The modality-specific encoders are aligned via contrastive learning.
\name extracts information from multimodal data using specialized encoders (Methods Sec.~\ref{method:framework}). 
\textsbf{c,} The missing data modality problem is evident in the scarcity of drugs with more than two available modalities. 
% While cell lines in transcriptomics are treated as separate modalities in the fusion module (Supplementary Note~\ref{note:architecture}; Supplementary Fig.~\ref{fig:supp_fig_1}c), throughout the text and illustrations, we refer to transcriptomics as a single modality to enhance readability and clarity. This distinction is intended to be self-evident from context. 
%
\textsbf{d,} UMAP of modality-specific latent embeddings of ten randomly sampled drugs, before and after modality alignment in \name. Prior to alignment, embeddings cluster by data type, while post-alignment, they cluster based on drug identity, enabling cross-modal integration. 
}
\label{fig:figure_1}
\end{figure}

Predicting effective combination therapies requires models that generalize across clinical and preclinical data modalities and predict accurately when some are missing, whereas many existing approaches assume complete modality coverage during training and inference and cannot be applied to novel compounds with limited clinical information. \name is a multimodal model designed to handle incomplete modality inputs during both training and inference. It predicts clinical outcomes and adverse reactions of drug combinations from preclinical data, supporting prioritization of both existing therapies and new candidates. Using structural drug information, a molecular pathways knowledge graph, transcriptomic responses, and cell viability data, \name predicts combination effects across 953 outcomes (for example, ``increase in QTc prolongation''; Fig.~\ref{fig:figure_1}a; Supplementary Note~\ref{note:outcomes}).

For each drug pair, \name encodes each modality with a modality-specific encoder and maps the resulting embeddings into a unified space through an attention bottleneck fusion module (Supplementary Note~\ref{note:architecture}). 
Bottleneck tokens are inserted between the transcriptomics embeddings and the structure, pathway, and cell viability embeddings to regulate information flow and to balance signal from transcriptomic responses across cell lines~\cite{Nagrani_2021_bottleneck}. 
The fusion module then integrates information from all available modality-specific embeddings via cross-attention and produces a single multimodal drug embedding~\cite{Recasens_2023_zorro, Jaegle_2022_perceiver_io}. Pairwise combination of the two drug embeddings followed by a prediction head yields a score for each outcome (Fig.~\ref{fig:figure_1}a,b; Supplementary Fig.~\ref{fig:supp_fig_1}{a}).
To align data modalities, \name uses contrastive learning that anchors modality-specific embeddings to the structure modality, which is universally available for small molecules (Fig.~\ref{fig:figure_1}a,b,c). After alignment, \name forms unified latent representations that enable fine-tuning on drug combination datasets (Methods Sec.~\ref{method:framework}; Fig.~\ref{fig:figure_1}d; Supplementary Fig.~\ref{fig:supp_fig_1}{a}). This design preserves predictive performance when some modalities are missing at inference and improves training efficiency.

\xhdrt{Drug combination benchmark}

\begin{figure}[t]
% \centering
\vspace*{-1cm}
\makebox[\linewidth]{
        \includegraphics[width=0.9\linewidth]{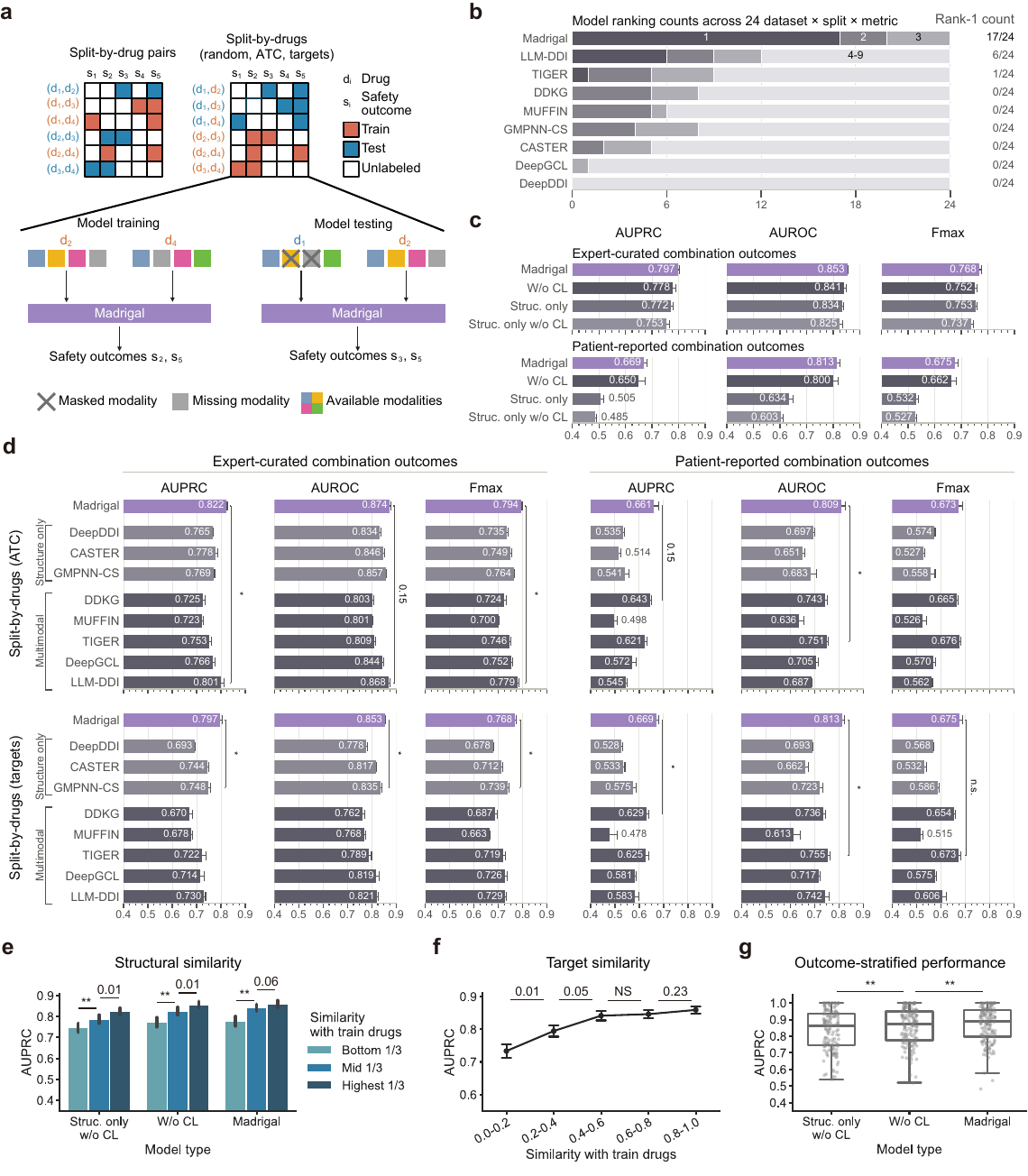}
    }
\caption{\textsbf{Benchmarking \name and performance analyses.} 
\textsbf{a,} Dataset splitting strategy for predicting safety outcomes of drug combinations. 
In the split-by-drugs setup, during training, all available modalities are used ($d_2, d_3, d_4$); at test time, test drugs are forced to only use the structure and transcriptomics modalities (patterned boxes) ($d_1$).
\revision{
\textsbf{b,} Rank counts of \name and existing methods across the 24 (dataset, split, metric) comparisons (2 datasets, 4 splits, and 3 metrics). 
\textsbf{c,} Ablation of \name components in the split-by-drugs (targets) setting. ``W/o CL", ablation model without modality alignment; ``Struc. only", ablation model with only structure modality during finetuning (all modalities available during modality alignment); ``Struc. only w/o CL", ablation model without modality alignment and with only structure modality available during finetuning. AUROC, area under the receiver operating characteristic curve; AUPRC, area under the precision-recall curve; Fmax, maximum of F-measure.
\textsbf{d,} Test performance of \name and existing methods in the split-by-drugs (targets) and split-by-drugs (ATC) settings in the DrugBank (expert-curated drug combination outcomes) and TWOSIDES (patient-reported drug combination outcomes) datasets. The brackets highlight Madrigal's improvements upon the best-performing existing method per (dataset, split, metric) comparison. **$p$-value$<0.005$, *$p$-value$<0.05$, exact value for $p$-value$<0.2$, n.s. otherwise; two-sided Mann-Whitney U test.
}
\textsbf{e,f,} Test performance increase for test drugs with increasing similarity to train drugs in terms of structure (e) or target profile (f). The progression from ``Struc. only w/o CL", ``W/o CL", to \name represents progressive additions of multimodal input and modality alignment upon a simple model only taking molecular structure as input. For each test drug, its structural similarity (with the train set) is calculated as the average of the highest 5 Tanimoto similarities between its Morgan fingerprint with any train drug's fingerprint. Target profile similarity is similarly defined as the average of the highest 5 Jaccard similarities between the drug's target profile with any train drug's profile. Target profiles of drugs are set of targets annotated to the drugs in DrugBank~\cite{drugbank}. Analyses in (e-g) are performed with models trained on the DrugBank dataset, split-by-drugs (random) setting. Error bars show 95\% confidence interval. Two-sided Mann-Whitney U test; **$p$-value $<0.005$. 
\textsbf{g,} Test performance of \name ablation with only structure modality and without modality alignment, ablation without modality alignment but with multi-modality, and full model, stratified by safety outcomes. Two-sided Wilcoxon signed rank test; **$p$-value $<0.005$.
}
\label{fig:figure_2}
\end{figure}

We evaluate \name on two settings: (1) holding out all samples for specific drug pairs (``split-by-drug pairs'') and (2) holding out all samples for specific drugs together with any of their pairings (``split-by-drugs''). The split-by-drugs setting better reflects prediction for a novel compound combined with an approved partner (Fig.~\ref{fig:figure_2}a).

\name is trained on two datasets: TWOSIDES, a FAERS-derived resource with 4{,}656{,}138 drug combination-outcome triplets across 1{,}457 drugs and 795 outcomes~\cite{twosides}, and DrugBank, an expert-curated resource with 1{,}188{,}371 drug combination-outcome triplets involving 3{,}632 drugs and 158 outcomes~\cite{drugbank} (Methods Sec.~\ref{method:data}). To probe \name's generalization, we introduce two harder variants of split-by-drugs. In split-by-drugs (targets), test drugs share minimal therapeutic targets with train drugs. In split-by-drugs (ATC), we exclude drugs of certain first-level Anatomical Therapeutic Chemical (ATC) categories from training (Methods Sec.~\ref{method:benchmark_dataset_split}). Both strategies increase structural dissimilarity between training and test drugs relative to random splits (Supplementary Fig.~\ref{fig:supp_fig_2a}{a}). For reference, example test drugs from the split-by-drugs (random) setting are shown alongside their closest training-set neighbors by Tanimoto similarity (Supplementary Fig.~\ref{fig:supp_fig_2a}{b}). For each dataset and split, models are trained separately and evaluated on the corresponding test set.

These splits mirror real development scenarios in which a novel compound is combined with an approved drug to mitigate safety issues, enhance efficacy, or extend indications despite limited preclinical data for the new agent. \revision{We compare against state-of-the-art models spanning two classes: structure-based models (DeepDDI~\cite{Kim_2023_deepddi_paxlovid}, CASTER~\cite{caster}, GMPNN-CS~\cite{nyamabo2022}); and multimodal models using structure, knowledge graph, interaction networks, and/or text (DDKG~\cite{Su_2022_ddkg}, MUFFIN~\cite{muffin}, TIGER~\cite{Su_2024_tiger}, DeepGCL~\cite{Jiang2024deepgcl}, LLM-DDI~\cite{Li2026llmddi}).} We report area under receiver-operator curve (AUROC), area under precision-recall curve (AUPRC), and maximum F measure (Fmax) (Methods Sec.~\ref{method:eval_metrics}).
To reflect realistic constraints when information is sparse for novel compounds, we restrict \name's test-time inputs to modalities typically available in early-stage drug development (Fig.~\ref{fig:figure_2}a). In contrast, other models receive their full multimodal inputs.

\xhdrt{Performance across hard generalization settings}

\revision{
\name achieves the best performance in 17 of 24 comparisons across datasets, splits, and metrics, significantly outperforming every other method in 14 of these comparisons (Fig.~\ref{fig:figure_2}b, Supplementary Fig.~\ref{fig:supp_fig_2b}{b}). 
Performance is consistently strong across both datasets under the most stringent generalization settings, split-by-drugs (targets) and split-by-drugs (ATC) (Fig.~\ref{fig:figure_2}d).
Under split-by-drugs (targets), \name achieves AUPRC 0.669$\pm$0.011, AUROC 0.813$\pm$0.012, and Fmax 0.675$\pm$0.013 on TWOSIDES. This is an average improvement of 11.9 percentage points over structure-based models across the three metrics and an average AUROC improvement of 10.0 percentage points over multimodal models. On DrugBank, \name reaches AUPRC 0.797$\pm$0.007, AUROC 0.853$\pm$0.004, and Fmax 0.768$\pm$0.007, with average gains of 5.7 percentage points over structure-based models across metrics and 6.1 percentage points over multimodal models in AUROC (Fig.~\ref{fig:figure_2}d).
% Because each positive sample is scored against two sampled negatives, an uninformative model attains an AUPRC of approximately 0.333; \name's AUPRC therefore corresponds to a 2.4-fold improvement over chance on DrugBank and 2.0-fold on TWOSIDES in this setting.
With two sampled negatives per positive sample, the chance-level AUPRC is approximately 0.333; \name therefore achieves approximately 2.0 and 2.4 times the chance-level AUPRC on TWOSIDES and DrugBank, respectively. Similar gains are observed under split-by-drugs (ATC) and split-by-drugs (random) (Fig.~\ref{fig:figure_2}d; Supplementary Fig.~\ref{fig:supp_fig_2b}{b}).

% Similar gains hold in split-by-drugs (ATC) and split-by-drugs (random), indicating that integrating and aligning modalities advances drug-combination outcome prediction (Fig.~\ref{fig:figure_2}d, Supplementary Fig.~\ref{fig:supp_fig_2b}{b}). 
%
The seven comparisons where \name does not rank first fall into two settings. In the split-by-drug pairs setting, where every test drug also appears in the training samples, LLM-DDI exceeds \name on all three metrics on both datasets. Under split-by-drugs (ATC), TIGER exceeds \name's Fmax on TWOSIDES by 0.003 (Supplementary Note~\ref{note:underperform}). 
% Three exceptions arise, two of them in the split-by-drug pairs setting, where every test drug appears in the training samples. TIGER exceeds \name's Fmax on TWOSIDES by 0.003 under split-by-drugs (ATC), and LLM-DDI and MUFFIN exceed \name in the split-by-drug pairs setting, LLM-DDI on DrugBank and TWOSIDES and MUFFIN on TWOSIDES (Supplementary Note~\ref{note:underperform}).
In a separate evaluation on outcome-stratified subsets of each test split, \name also outperforms GPT-5.4 in all 24 comparisons. Its AUPRC exceeds that of GPT-5.4 by 11.8--23.8 percentage points on DrugBank and by 20.8--32.7 percentage points on TWOSIDES (Supplementary Table~\ref{tab:gpt54_subset}; Supplementary Note~\ref{note:gpt54}).
}

We next examine model robustness. 
% We hypothesize that test drugs with greater representation and higher structural similarity to train drugs will yield higher accuracy. 
Accuracy increases with the structural similarity of test drugs to training drugs for the full model, the model without contrastive alignment (w/o CL), and the structure-only model without contrastive alignment (Struc. only w/o CL) (Fig.~\ref{fig:figure_2}e). Higher similarity in target profiles is also associated with better performance (Fig.~\ref{fig:figure_2}f). Predictions are more accurate for narrowly defined outcomes that map to specific biological pathways than for broader phenotypes, consistent with the value of pathway-resolved knowledge (Supplementary Fig.~\ref{fig:supp_fig_2a}{e})~\cite{Zitnik2018}.

Ablation experiments further show that both multimodal input and contrastive modality alignment contribute to performance gains across all datasets, splits, and metrics (Fig.~\ref{fig:figure_2}c,g; Supplementary Fig.~\ref{fig:supp_fig_2b}{a}). 
% \name outperforms its unimodal counterparts, and the gains are largest when modalities are aligned (Fig.~\ref{fig:figure_2}g). 
\revision{Performance improves as additional modalities are included. Pathway knowledge contributes most, and experimental modalities provide predictive signal for compounds without pathway annotations (Supplementary Fig.~\ref{fig:supp_fig_3}{a,d}; Supplementary Note~\ref{note:modality_attention}). Adding a bioassay modality yields further gains across datasets and splits (Supplementary Table~\ref{tab:bioassay_results}; Supplementary Note~\ref{note:bioassay}). Attention analyses show substantial attention to transcriptomic tokens despite the low availability of transcriptomics data across compounds (Supplementary Fig.~\ref{fig:supp_fig_3}{b,c}; Supplementary Note~\ref{note:modality_attention}).}

\xhdrt{Predicting single-drug toxicity and transporter-mediated drug-drug interactions}

\begin{figure}[t]
% \centering
% \vspace*{-2cm}
\makebox[\linewidth]{
        \includegraphics[width=1\linewidth]{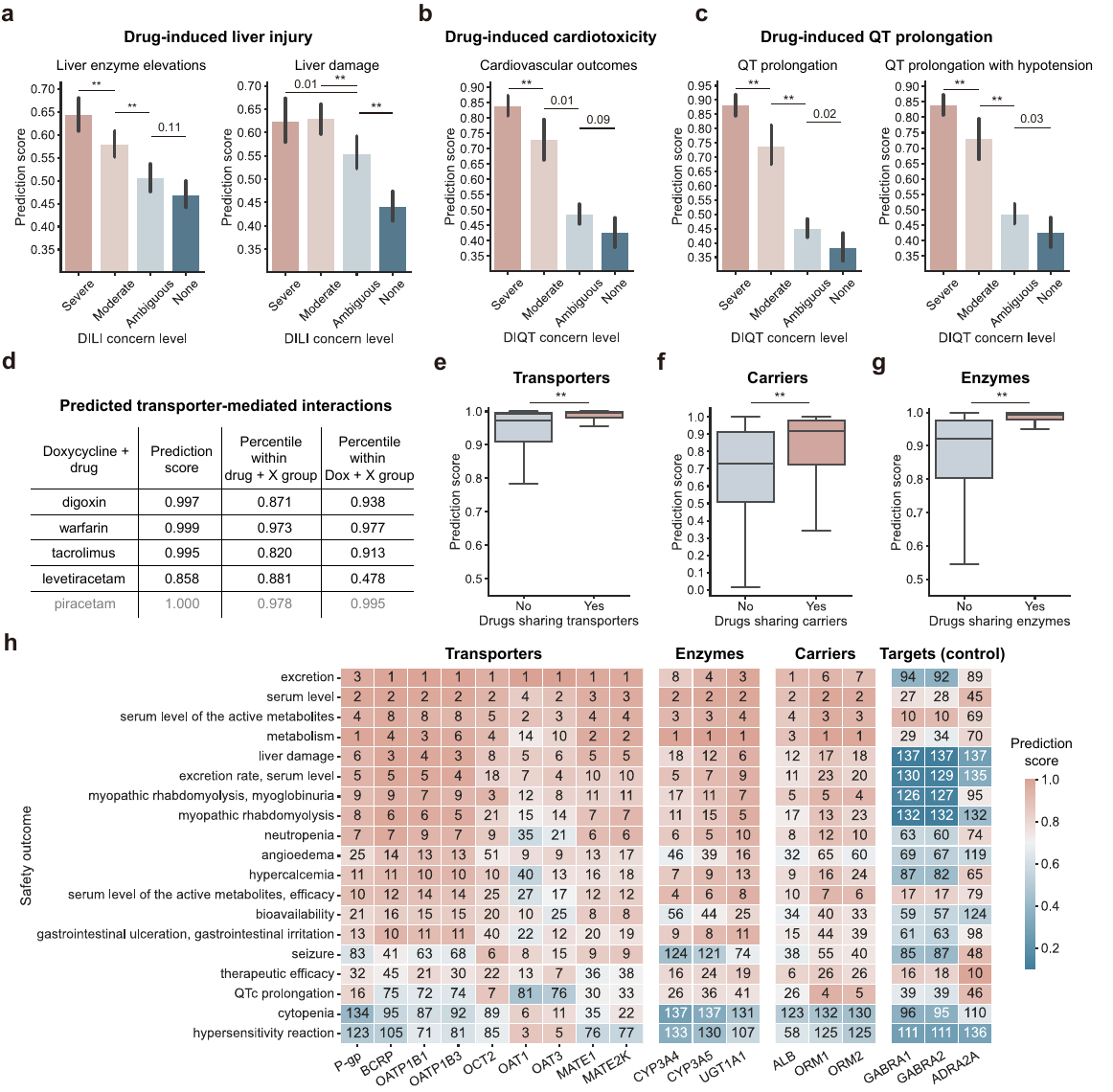}
    }
\caption{\textsbf{Evaluating \name predictions on external patient safety datasets.} 
\textsbf{a-c,} Model predictions of individual drug's organ-specific adverse effects correlate with concern levels in three organ-specific adverse effect datasets (drug-induced liver injury (a), drug-induced cardiotoxicity (b), drug-induced QT prolongation (c)). 
Error bars show 95\% confidence interval. Two-sided Mann-Whitney U test; ** $p$-value $< 0.005$. Higher prediction score indicates greater predicted concern.
\textsbf{d,} Model predictions of transporter-mediated DDIs (Methods Sec.~\ref{method:transportome}) in combinations involving doxycycline (Dox). Piracetam is included as a reference as it is structurally highly similar to levetiracetam. For each drug pair, the prediction score column denotes the maximal prediction score among transporter-mediated safety outcomes that are relevant to the increase in serum concentration from DrugBank. Percentiles compare the max prediction score of Dox + X among either X + any curated DrugBank drug (``drug + X group") or Dox + any curated DrugBank drug (``Dox + X group").
\textsbf{e-g,} Drugs sharing the same transporters (e), carriers (f), or enzymes (g) are predicted to have a higher tendency to have relevant safety outcomes (Methods Sec.~\ref{method:transportome}). Transporter, carrier, and enzyme information of drugs are obtained from DrugBank~\cite{drugbank}. The highest prediction score among all potential transporter-, carrier-, or enzyme-mediated safety outcomes is considered for each drug pair (Methods Sec.~\ref{method:transportome}). Two-sided Mann-Whitney U test; **$p$-value $< 0.005$.
\textsbf{h,} Drugs sharing specific transporters are predicted to have a higher tendency of both common and specific transporter-related safety outcomes. 
Safety outcomes shown are ranked in the highest 10 for at least one transporter (across all drug pairs sharing it). The color gradient reflects the median prediction score (across all drug pairs sharing corresponding transporter), and the number in each cell is the ranking (max=141) of the median prediction score of the corresponding safety outcome among all safety outcomes for drug pairs sharing the corresponding transporter. Outcome signs are neutralized before ranking, collapsing the 158 DrugBank safety outcomes into 141 (Methods Sec.~\ref{method:transportome}). The safety profiles of drugs sharing three enzymes, carriers, and targets, respectively, are also shown for comparison. }
\label{fig:figure_3}
\end{figure}

To assess whether \name captures safety signals beyond combination contexts, we evaluate it on individual drugs by pairing each drug with itself. \name's predictions correlate with established safety profiles for liver injury (DILIrank)~\cite{Chen_2016_DILI}, cardiotoxicity (DICTrank)~\cite{Qu_2023_DICT}, and QT prolongation (DIQTA)~\cite{Li_2022_DIQTA} (Fig.~\ref{fig:figure_3}a-c; Supplementary Fig.~\ref{fig:supp_fig_4}), indicating that \name surfaces clinically relevant single-drug risks from preclinical modalities. \revision{Supplementary Table~\ref{tab:analysis_overview} summarizes the question, inputs, outputs, comparison basis and interpretation.}

We next ask whether \name's predictions reflect patterns consistent with transporter-mediated interactions~\cite{Shi_2024_transportome, Galetin_2024_transporter_organ}. Membrane transporters influence absorption, distribution, and elimination, and two drugs that share the same transporter can compete for it and produce a clinically meaningful interaction. The International Transporter Consortium reported that about 75\% of the top 200 prescribed drugs are substrates of at least one transporter, and many engage several transporters, which raises their potential for transporter-mediated DDIs~\cite{international2010membrane}. \name predicts elevated risk for newly identified transporter-mediated DDIs, including doxycycline with digoxin, warfarin, tacrolimus, and levetiracetam (Fig.~\ref{fig:figure_3}d). Doxycycline + tacrolimus receives a high predicted score, and doxycycline + levetiracetam, a pair absent from training, receives an elevated rank among levetiracetam's combinations.

\name can prioritize potential transporter-mediated risks of interaction in novel combinations. Drugs that share transporters, enzymes, or carriers receive significantly higher predicted scores for the corresponding safety outcomes (Fig.~\ref{fig:figure_3}e-g). For drugs sharing specific transporters that regulatory guidance prioritizes because of organ-specific safety risks~\cite{Galetin_2024_transporter_organ}, \name predicts elevated scores for the corresponding transporter-related safety events (Fig.~\ref{fig:figure_3}h).

\xhdrt{Clinical trial concordance for combination safety}

\begin{figure}[t]
% \centering
% \vspace*{-2cm}
\makebox[\linewidth]{
        \includegraphics[width=1\linewidth]{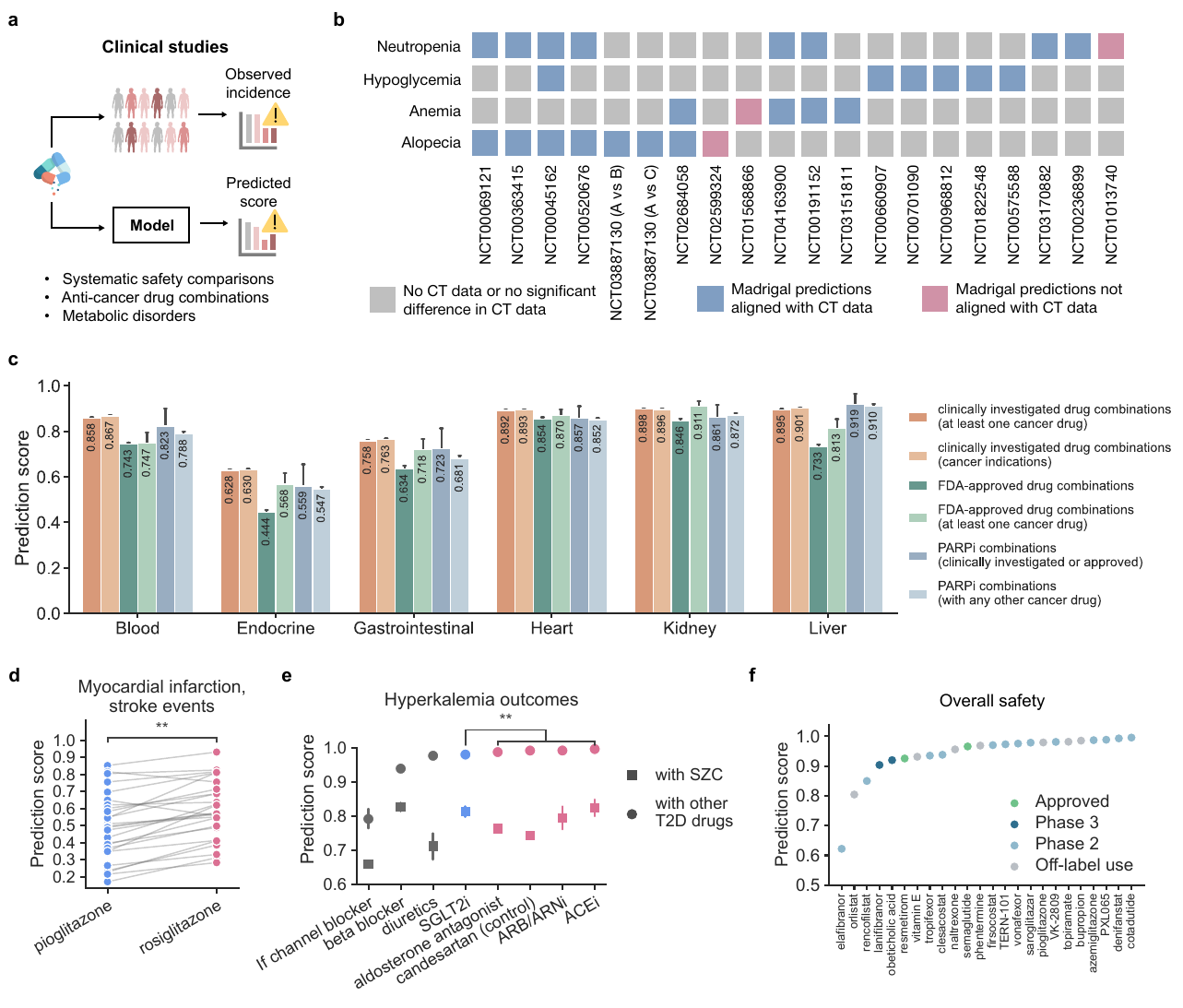}
    }
\caption{\textsbf{\name predicts safety for clinically tested drug combinations.} 
\textsbf{a,} A candidate drug pair is evaluated in two ways. Top: head-to-head clinical trial arms yielding observed AE incidence. Bottom: \name predicts safety scores with regard to the same adverse outcome. The model predictions are not calibrated to match the percentage. Agreement is assessed by whether the safer arm in the trial also receives a lower \name score. 
\textsbf{b,} Comparing \name predictions with clinical trials (CT) AE data in advanced-stage clinical trials with multiple combination arms for neutropenia, hypoglycemia, anemia, and alopecia. 
\textsbf{c,} Comparative safety assessment across different classes of drug combinations. 
Left to right bars within each group represent: (1) drug combinations containing at least one cancer drug that have been investigated in advanced stage (above Phase I), (2) drug combinations indicated for cancer that have been investigated in advanced stage, (3) FDA-approved drug combinations, (4) FDA-approved drug combinations containing at least one cancer drug, (5) PARPi combinations that have been investigated in advanced stage, and (6) pairwise combinations of a PARPi with any other cancer drug. 
Each drug combination's safety profile is represented by the average of the five highest prediction scores for outcomes of each organ system. Higher prediction score indicates greater predicted concern.
% \revision{
\textsbf{d,} Predicted safety profiles of combination involving pioglitazone or rosiglitazone with T2D drugs. 
Each point represents the mean prediction score of pioglitazone or rosiglitazone, when combined with any T2D drug of a different mechanism of action, with regard to each relevant safety outcome. 
One-sided Wilcoxon signed rank test; **$p$-value $< 0.005$.
\textsbf{e,} Predicted hyperkalemia-related safety profiles of drug combination involving a heart failure drug and any T2D drug of a different mechanism of action. 
SGLT2i, sodium/glucose cotransporter 2 inhibitor; ARB, angiotensin II receptor blocker; ARNi, angiotensin receptor/neprilysin inhibitor; ACEi, angiotensin-converting enzyme inhibitor; SZC, sodium zirconium cyclosilicate; HF, heart failure. Error bars show 95\% confidence interval. One-sided Mann-Whitney U test; **$p$-value $< 0.005$.
\revision{
\textsbf{f,} Predicted safety of MASH clinical candidates from~\cite{Harrison_2023_clinical_landscape_nash} in combination with T2D drugs, calculated from the average of highest 5 prediction scores across outcomes (Methods Sec.~\ref{method:t2d_comorb}). Shown all MASH candidates, ranked by prediction scores. 
}
}
\label{fig:figure_5}
\end{figure}

Randomized head-to-head trials of combination regimens provide a controlled setting to test if predicted safety orderings align with observed arm-level adverse-event outcomes. 
We identify 35 advanced-stage trials since 2000 that tested multiple small-molecule combinations under the same conditions (Methods Sec.~\ref{method:clinical_trials_data}; Supplementary Table~\ref{tab:trial_arms}). For each trial, we select adverse events (AEs) with significantly different incidences between arms (Methods Sec.~\ref{method:clinical_trials}). We then performed a directional agreement check, where agreement is defined as the trial arm with a more favorable safety profile also receiving a lower \name score (Methods Sec.~\ref{method:clinical_trials}). Across AEs with at least five arm comparisons, \name's risk ordering agrees with trial outcomes in 7/8 arm pairs for alopecia AE, 4/5 arm pairs for anemia AE, 6/6 arm pairs for hypoglycemia AE, and 8/9 arm pairs for neutropenia AE (Fig.~\ref{fig:figure_5}a,b; Supplementary Tables~\ref{tab:trial_odds_alopecia}-\ref{tab:trial_odds_neutropenia}, 19 trials with significantly different incidences between arms for at least one of neutropenia, hypoglycemia, anemia, and alopecia). 

We next test whether \name differentiates combinations that have progressed into the clinic. \name recapitulates known hematologic toxicities of poly(ADP-ribose) polymerase inhibitor (PARPi) combinations, such as the higher rates of grade 3-4 hematologic AEs, notably neutropenia, reported for (olaparib + paclitaxel) in gastric cancer~\cite{Bang_2017_olaparib_paclitaxel}, which has so far not been approved (Supplementary Fig.~\ref{fig:supp_fig_5}{a-c}; Supplementary Note~\ref{note:parpi_safety}). 

PARPi combinations under investigation or approval~\cite{Bhamidipati_2023_parpi_clinical_combos,DailyMed_LYNPARZA_olaparib,DailyMed_TALZENNA_talazoparib} are predicted to have safety profiles comparable to pairing PARPi with any other cancer drug across all six organ systems, differing by at most 0.042 in predicted score and falling within the confidence intervals of the estimates (Fig.~\ref{fig:figure_5}c; Supplementary Note~\ref{note:cancer_combos}). In all organs except liver, these PARPi combinations are also predicted to be safer than clinically investigated oncology combinations, including those active in 2024, failed, or withdrawn~\cite{Shtar_2022_CDCDB}. Combinations already used in patients (US FDA Orange Book) are predicted to be safest overall across organs, with cardiovascular outcomes the second safest organ system.
We visualize prediction scores across outcomes for each PARPi combination under clinical investigation. PARPi combinations that have advanced further clinically or are approved generally receive more favorable safety predictions than those in earlier phases (Supplementary Fig.~\ref{fig:supp_fig_5}{d}; ordered left to right by increasing average of the top five highest prediction scores across all safety outcomes).

\xhdrt{Polypharmacy case studies in T2D and MASH}

The management of chronic metabolic disorders often requires complex polypharmacy due to multimorbidity and multifactorial pathophysiology~\cite{Langenberg_2023_multimorbidity}. This is pronounced in type 2 diabetes (T2D) and metabolic dysfunction-associated steatohepatitis (MASH), which are increasingly prevalent~\cite{Ong_2023_t2d_burden,Younossi_2018_nash_review} and heterogeneous in mechanism~\cite{DelPrato_2019_t2d_combo,Ratziu_2023_rational_combo_nash,Tilg_2023_nash_challenges,Harrison_2023_nash_challenges}. MASH frequently co-occurs with T2D~\cite{Younossi_2018_nash_review}, with global prevalence estimates rising from 5-7\% in the general population to 37\% among people with T2D~\cite{Younossi_2024_nash_burden}. The first MASH therapy, resmetirom, was approved in 2024~\cite{nash_approval}. We assess \name's utility in three polypharmacy settings: (1) T2D and heart failure, a well-characterized comorbidity; (2) T2D and MASH, an emerging area; and (3) MASH combination therapy, where options are limited.

\underline{T2D and heart failure.} \name's safety rankings align with clinical knowledge. Combinations involving rosiglitazone are predicted to have less favorable cardiovascular profiles than those with pioglitazone, consistent with reports of myocardial infarction and stroke risk~\cite{Nissen_2007_rosiglitazone} (Fig.~\ref{fig:figure_5}d; Supplementary Table~\ref{tab:t2d_drugs}, \ref{tab:mi_stroke_outcomes}). When pairing heart-failure therapies with glucose-lowering drugs, \name reflects the hyperkalemia risk associated with renin-angiotensin-aldosterone system inhibitors~\cite{Weinstein_2021_hyperkalemia} and the mitigating association of SGLT2 inhibitors~\cite{Neuen_2022_sglt2i_hyperkalemia} (Supplementary Table~\ref{tab:hf_drugs}). Combinations including sodium zirconium cyclosilicate, used to treat hyperkalemia, are predicted to have improved safety relative to other pairings, consistent with recent clinical practice and trial design~\cite{Oshima2021szc,Imamura2021szc,Kosiborod2024realizek}; candesartan, with known hyperkalemia risk~\cite{Desai_2007_charm_cardesartan}, serves as a control (Fig.~\ref{fig:figure_5}e). For renal effects, combinations with SGLT2 inhibitors are predicted to have more favorable renal safety than those with diuretics, in line with clinical observations~\cite{Heerspink_2020_sglt2i_kidney,Lin_2024_diuretic_sglt2i_kidney} (Supplementary Fig.~\ref{fig:supp_fig_6}{a}).

\underline{T2D and MASH.} We evaluate safety profiles when MASH drugs or candidates (approved, in trials, or used off-label~\cite{Harrison_2023_clinical_landscape_nash}) are combined with T2D drugs of different mechanisms (Fig.~\ref{fig:figure_5}f; Supplementary Table~\ref{tab:mash_drugs}; Methods Sec.~\ref{method:t2d_comorb}). \name ranks resmetirom, the first FDA-approved MASH drug, as the sixth most favorable, aligning with its known favorable safety profile. All Phase 3 candidates are also ranked in top five. The top-ranked candidate, elafibranor, has shown a consistent safety profile across trials, including in primary biliary cholangitis~\cite{Kowdley_2024_elafibranor}; although the RESOLVE-IT Phase III trial (NCT02704403) did not meet its primary MASH efficacy endpoint, safety and tolerability were consistent with prior studies, suggesting a potential role for elafibranor within combination regimens. Some risks are not captured by our label sets: tropifexor's dose-related pruritus in Phase II~\cite{Sanyal_2023_tropifexor} and firsocostat-associated hypertriglyceride\-mia~\cite{Alkhouri_2020_firsocostat_hypertriglyceridemia} are not annotated in our DrugBank-derived outcome set, which may lead to underestimation in our predictions. Across candidates, combinations involving Phase I candidates, where clinical programs focus on safety, tend to rank less favorably than those in later phases (Supplementary Fig.~\ref{fig:supp_fig_6}{d}), and safety varies by mechanism of action when paired with T2D drugs (Supplementary Fig.~\ref{fig:supp_fig_6}{b}).

\underline{MASH combination strategies.} MASH combinations are typically motivated by either improved efficacy (targeting independent pathways or multiple nodes of one pathway) or improved safety (a second agent mitigates the first agent's adverse effects)~\cite{Ratziu_2023_rational_combo_nash} (Supplementary Fig.~\ref{fig:supp_fig_6}{e}). We annotate combinations under clinical investigation~\cite{Suri_2022_nash_combos,Harrison_2023_nash_challenges,Tilg_2023_nash_challenges,Ratziu_2023_rational_combo_nash} and rank their safety using the average of the top five prediction scores across outcomes. Each regimen is compared against background combinations formed from other pairings between drug pairs with the same respective mechanisms of action (Methods Sec.~\ref{method:nash_combo}). 
By this criterion, 3/5 combinations curated for enhanced safety and 3/11 curated for enhanced efficacy rank in the safer half of their background combinations (Supplementary Fig.~\ref{fig:supp_fig_6}{f}).

\xhdrt{Prediction of efficacy in patient-derived models}

\begin{figure}[t]
\makebox[\linewidth]{
        \includegraphics[width=1\linewidth]{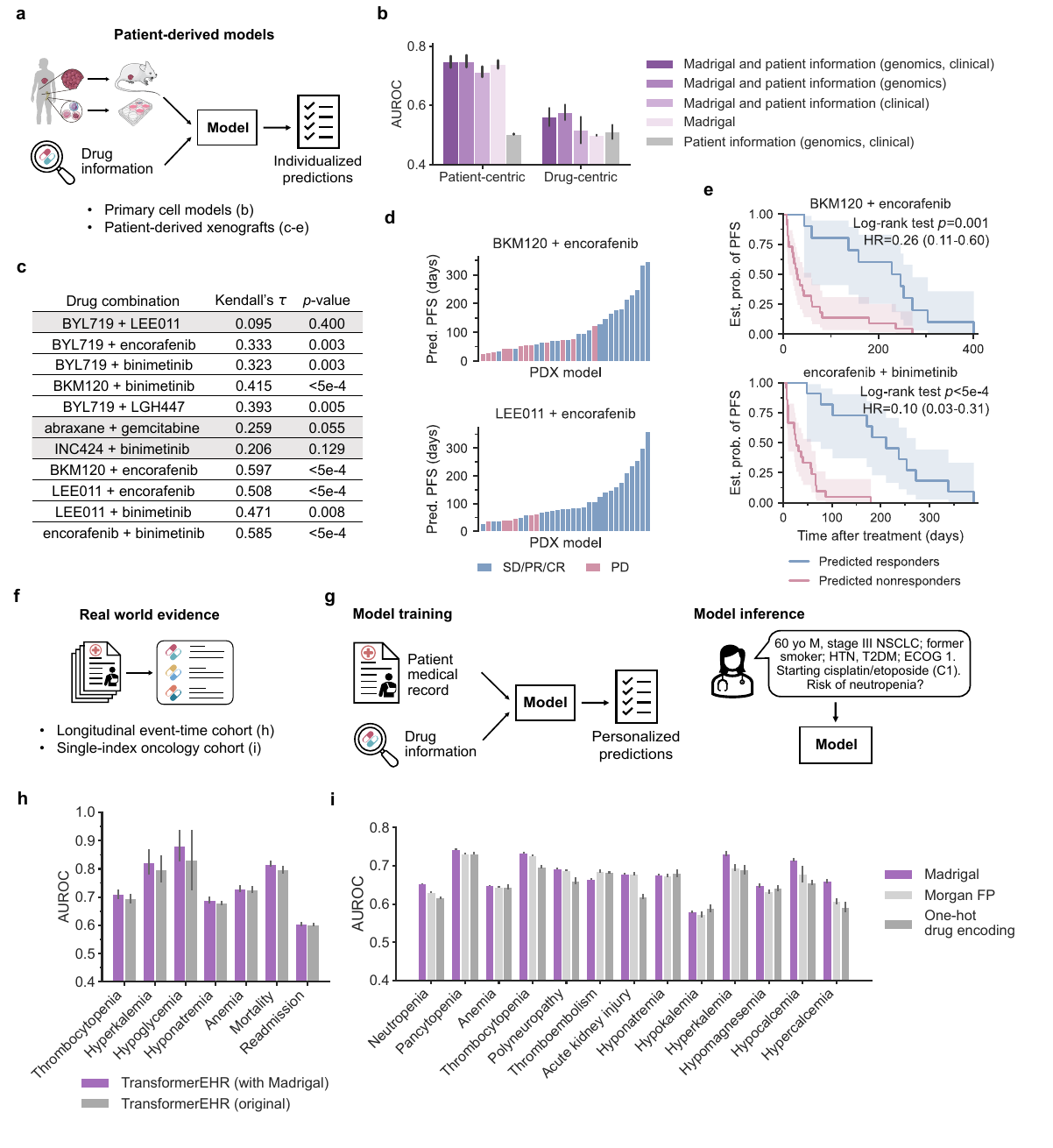}
     }
\caption{\textsbf{\name predicts individualized drug combination efficacy in patient-derived cancer models and adverse events in real-world patient cohorts.} 
\textsbf{a,} Using \name to predict individualized drug combination efficacy in patient-derived cancer models.
\textsbf{b,} Performance of \name in predicting synergistic drug combinations in BeatAML~\cite{Bottomly_2022_BeatAML}, where prediction target is combination synergy (Methods Sec.~\ref{method:beataml}). 
Model evaluation is conducted on randomly held-out patients. 
Patient-centric and drug-centric denote two ways of calculating AUROC to evaluate the model (Methods Sec.~\ref{method:beataml}). Error bars show standard deviation across 25 runs (five \name models $\times$ five training seeds). 
\textsbf{c,} Performance of \name in predicting drug combination efficacy in PDX Encyclopedia\cite{Gao_2015_pdx} when leaving each drug combination out. The prediction target is treatment response (BestAvgResponse).
\revision{
\textsbf{d,} Predicted progression-free survival (PFS, TimeToDouble) for individual patient models treated with the (BKM120 + encorafenib) and (LEE011 + encorafenib) combinations. The predictor is trained on other drug combinations, with PFS as the prediction target. Predictions are color-coded by the observed best response category (calculated from response according to mRECIST~\cite{Gao_2015_pdx}) of each patient model. PD, progressive disease; SD, stable disease; PR, partial response; CR, complete response.
\textsbf{e,} Kaplan-Meier survival estimates stratified by predicted treatment response for the (BKM120 + encorafenib) and (encorafenib + binimetinib) combinations. The predictor is trained on other drug combinations with treatment response as the prediction target (same predictor as in (c)). 
}
%
% \revision{
\textsbf{f,} Using \name to predict personalized drug combination adverse events in real-world patient cohorts.
\textsbf{g,} Model training and inference for predicting drug combination outcomes in patients.
\textsbf{h,} Performance of TransformerEHR and TransformerEHR with \name drug embeddings across re-admission prediction, mortality prediction, and adverse event prediction (anemia, hypoglycemia, hyperkalemia, hyponatremia, thrombocytopenia) tasks in the longitudinal event-time cohort. Error bars show standard deviation.
% Each result was averaged from five runs. 
\textsbf{i,} Performance of combining \name with patient information to predict adverse events for individual patients in the independent oncology cohort, compared with using Morgan fingerprint or one-hot regimen encoding. Error bars show standard deviation.
}
\label{fig:figure_7}
\end{figure}

We test whether \name can support individualized prioritization of cancer drug combinations~\cite{Sicklick_2019_ipredict,Meric_2023_combomatch} by pairing its drug embeddings with patient molecular profiles in two patient-derived systems: primary cancer cells from BeatAML and patient-derived xenografts from PDXE (Fig.~\ref{fig:figure_7}a; Supplementary Fig.~\ref{fig:supp_fig_7}a).

\underline{BeatAML.} For each drug combination, \name's two drug embeddings are fused with a bilinear decoder. The fused embedding is concatenated with dimensionality-reduced gene expression, somatic mutation profiles, and clinical attributes, then passed to a multi-layer perceptron (Methods Sec.~\ref{method:beataml}). The classification target is synergy defined by a drug combination ratio less than 1~\cite{Eide_2023_Venetoclax}. Models that combine \name with patient genomic profiles outperform models without genomics, improving drug-centric AUROC by $0.06$ when patients are held out (Fig.~\ref{fig:figure_7}b) and by $0.09$ when drugs are held out (Supplementary Fig.~\ref{fig:supp_fig_7}b) over \name embeddings alone, indicating the predictive value of patient genomic information. Patient features alone are insufficient without the combined drug embeddings (AUROC $\approx0.50$).

\underline{PDXE.} We integrate gene expression and mutation data with \name and evaluate performance by holding out each drug combination. The element-wise maximum of the two \name drug embeddings is concatenated with dimensionality-reduced gene expression and mutation data and used as input to a random forest regressor \revision{(results are robust to the reduction method; Supplementary Note~\ref{note:pdx_robustness})}. We train two predictors: treatment response (BestAvgResponse) and progression-free survival (PFS, TimeToDouble)~\cite{Gao_2015_pdx}. For response, \name shows significant correlations between predicted and observed responses in 8 of 11 held-out combinations (Kendall's $\tau$, $p$-value$<0.05$), with an average $\tau=0.453$ among the eight (Fig.~\ref{fig:figure_7}c). We stratify patients using an mRECIST-based threshold (predicted response below $-20$ as responders, otherwise non-responders)~\cite{Gao_2015_pdx}. For combinations with at least five predicted responders and five predicted non-responders, Kaplan-Meier estimates show significant differences in ground-truth PFS between strata in four of the five such combinations (log-rank $p<0.05$; Fig.~\ref{fig:figure_7}e; Supplementary Fig.~\ref{fig:supp_fig_7}c). For PFS prediction, \name again correlates with observed survival in 7 of 11 combinations ($p$-value$<0.05$, average $\tau=0.491$; Supplementary Fig.~\ref{fig:supp_fig_7}d). Per-model predicted PFS also tracks the observed best response category (Fig.~\ref{fig:figure_7}d; Supplementary Fig.~\ref{fig:supp_fig_7}f). These results show that \name can transfer from multimodal preclinical drug information to patient-level efficacy when combined with genomic context.

\xhdrt{Adverse-event prediction in longitudinal healthcare and oncology cohorts}

Having established that \name supports individualized combination response in patient-derived models, we next evaluate adverse-event prediction in patient cohorts (Fig.~\ref{fig:figure_7}f,g).

\underline{Longitudinal healthcare cohort.} We integrate \name's drug combination embeddings into TransformerEHR~\cite{Yang2023transformerehr} and fine-tune the extended TransformerEHR on longitudinal health records from Stanford Medicine (n=6,739)~\cite{ehrshot} (Methods Sec.~\ref{method:ehrshot_data}). We consider medication-related outcomes: hospital re-admission, all-cause mortality, and five AEs (thrombocytopenia, hyperkalemia, hypoglycemia, hyponatremia, anemia) (Methods Sec.~\ref{method:ehrshot_task}). Using \name with TransformerEHR reaches an average AUROC of 0.75 across all seven outcomes, improving consistently over the original TransformerEHR in all seven outcomes and by 1.8 percentage points in AUROC on average (Wilcoxon signed-rank $p=0.016$; Fig.~\ref{fig:figure_7}h; Supplementary Table~\ref{tab:ehrshot_perf}).

\underline{Oncology cohort.} We curate a Dana-Farber Cancer Institute cohort of patients treated with first-line regimens comprising exactly two small-molecule oncology drugs between June 2015 and March 2025 (n=3{,}577; 26 two-drug regimens). We track 13 AEs across five classes: hematotoxicity, neuropathy, thromboembolism, renal impairment, and fluid/electrolyte imbalance (Supplementary Table~\ref{tab:ae_icd_mapping}; Methods Sec.~\ref{method:dfci_data}). 

At the population level, we quantify the association between \name-predicted scores and observed AE incidence using Kendall's $\tau$ (Methods Sec.~\ref{method:dfci_task}). \name scores correlate significantly with AE incidence for 9 of 13 AEs (Kendall's $\tau=0.43\text{-}0.68$, $p$-value$<0.05$; Supplementary Table~\ref{tab:dfci_kendall_tau}). 
In multivariable logistic models adjusted for age, gender, race, and palliative intent, \name scores are independent predictors for 9 of 13 AEs with log-odds coefficients of 0.46-1.41 per standard deviation of score (Wald $p<0.05$; Supplementary Table~\ref{tab:dfci_log_odds}).
These associations hold for events with low incidence in the cohort: among the nine AEs with significant Kendall correlations, maximum incidence across treatment regimens ranges from 4\% (hyperkalemia) to 14\% (neutropenia), suggesting that \name's scores track observed risk even for infrequent events.
All 13 coefficients are positive, indicating that higher \name scores are consistently associated with increased AE incidence. 

We then evaluate patient-level prediction by combining \name drug embeddings with clinical covariates (age, gender, palliative intent, and race) in random forest models, and assess performance by five-fold stratified cross-validation. Using \name embeddings achieves an average AUROC of 0.68 and outperform both Morgan structural fingerprints and one-hot drug identity encodings in 9 of 13 outcomes (11 of 13 versus Morgan and 10 of 13 versus one-hot individually), with a significant per-outcome advantage (two-sided Mann-Whitney U test, $p$-value$<0.05$) over Morgan fingerprints in 9 of 13 outcomes and over one-hot encodings in 8 of 13 (Fig.~\ref{fig:figure_7}i; Supplementary Table~\ref{tab:dfci_auroc}). These gains suggest that multimodal drug representations can contribute predictive signal to the prediction of patient-level adverse-event outcomes beyond structural fingerprints and drug-identity indicators.

\section*{Discussion}
Combination therapies are central to modern care, yet translating preclinical evidence into clinical combination safety is difficult. 
We develop \name, a multimodal AI model that predicts the effects of drug combinations from preclinical modalities, molecular structure, target and pathway annotations, cell-viability profiles, and transcriptomic responses to perturbation. \name encodes each modality separately, aligns the representations in a shared latent space by contrastive learning, and fuses the representations a drug has into one embedding, so that a compound characterized in one or two modalities still receives a prediction. Alignment spans 21,842 approved and novel compounds, and prediction covers 158 expert-curated and 795 patient-reported clinical outcomes. \name surpasses single-modality and multimodal models for adverse drug interaction prediction, recovers transporter-mediated interactions observed clinically, and tracks the safety of clinically tested combinations and of combinations used in T2D and MASH. \revision{\name also predicts individualized outcomes from BeatAML genomic profiles, patient-derived xenografts, and real-world patient data.} 

\name scores combinations that have entered preclinical evaluation without clinical safety data. ComboMATCH is an NCI platform trial that assigns combination regimens on the basis of molecular alterations, and its candidate regimens have no such data. \name predicts safety risks for eight of these regimens (Supplementary Fig.~\ref{fig:supp_fig_discuss}c, Supplementary Note~\ref{note:combomatch})~\cite{Meric_2023_combomatch}. We also compare \name predictions against a proteome-wide atlas of drug mechanism of action~\cite{Mitchell_2023_proteome_moa}, which measures protein abundance after perturbation by each of hundreds of compounds. Drugs that induce similar proteomic changes receive similar \name embeddings and higher predicted interaction scores, and this association holds after adjusting for similarity of annotated target profiles (Supplementary Fig.~\ref{fig:supp_fig_discuss}a,b, Supplementary Note~\ref{note:proteome}). 

\revision{Multimodal models such as \name can predict outcomes for combinations before they are tested, and they impose no fixed requirement on how drugs are represented. This flexibility is important because drug-combination effect datasets are incomplete. Outcomes reported in the clinical literature are often absent from the annotation sets used for training~\cite{dent2013phase,dent2010safety}, and prescribing patterns shape which combinations appear at all. Clinicians prescribe SGLT2 inhibitors more frequently in T2D with heart failure~\cite{vaduganathan2022sglt2}, so the exposed populations differ in baseline risk from those exposed to alternatives, and the recorded event rates reflect that difference. Linking \name to an LLM interface addresses the first problem. A user states a clinical effect in free text, the LLM maps it onto the outcome space the model was trained on, and \name scores candidate combinations against that description before experimental testing (Supplementary Fig.~\ref{fig:llm_supp_fig}, Supplementary Notes~\ref{note:llm_extension} and~\ref{note:llm_impl}).}

\revision{Our evaluation is retrospective. The findings on clinical alignment, mechanistic interpretation, and personalized prediction are hypothesis-generating and not clinically actionable evidence, and prospective evaluation in independent cohorts will be required before \name can guide combination selection in practice.} Three properties of the training data bound what the predictions can mean. First, the labels inherit the reporting and prescribing practices that generated them, so a predicted risk partly tracks how often a combination has been observed and reported instead of the hazard intrinsic to the pair. 
Second, the datasets record an outcome for a drug pair without the indication, the population, or the clinical context in which it occurred, so \name screens candidate pairs across contexts instead of assessing a pair within a defined indication.
Firsocostat illustrates the consequence. It is an OATP1B1/1B3 substrate, and these transporter activities decline with cirrhosis~\cite{Younis_2024_firsocostat_oatp1b1}, so its interaction risk with other potent OATP1B1/1B3 substrates depends on liver function in a way the training labels do not capture. \revision{Third, modality coverage is sparse and uneven, so a prediction for a compound observed in few modalities draws largely on structure and rests on less evidence than the aggregate metrics suggest.} Accuracy also drops for underrepresented drug classes even when a specific drug has rich preclinical data. Adavosertib has cell-viability and transcriptomic data, but it is the only WEE1 inhibitor in the dataset, so the model sees no class-consistent perturbation pattern to generalize from, which may limit detection of WEE1-linked hematologic toxicities~\cite{Zhang2024wee1,Tutt2022violette}. \revision{A systematic error analysis identifies \name's potential blind spots, mainly among nervous-system drugs, organic acids and organic cation transporter substrates with hematological, metabolism and cardiovascular/renal outcomes (Supplementary Note~\ref{note:error_analysis}). Our modality-contribution analyses are correlational, since attention weights indicate which modality tokens the readout draws on, not which mechanism produces an outcome.}

\revision{A second set of limitations follows from how \name formulates the prediction task. It treats each outcome as a binary label attached to a drug pair, without severity grading and without a time axis, so it does not separate an event that appears within days from one that accumulates over months of exposure. Late-onset myelosuppression, chronic organ toxicity, and efficacy decay during treatment interruptions and dose adjustments therefore fall outside the model.} Clinical failures of drug development arise primarily from insufficient efficacy (40-50\%), followed by toxicity (20-30\%) and pharmacokinetic issues (10-15\%)~\cite{Harrison2016clinicalfailure,Dowden2019clinicalsuccess}, so a favorable \name safety screen is necessary but not sufficient to de-risk a combination. \revision{Deployment would further require predictions resolved by outcome severity and not by incidence alone, and reported with calibrated uncertainty, so that a user can distinguish a confident low-risk prediction from an uninformative one.}

One direction for future work is to predict efficacy and safety together. Coupling \name with efficacy models and PK/PD simulators that connect exposure to target engagement would support that, and explicit dose modeling would as well. Dose is currently implicit in the sources \name learns from, because FAERS reports, drug labels, and trial reports describe outcomes at whatever dose was administered without exposing that dose to the model. Harmonized PK parameters, together with physiologically based PK modeling, would make dose explicit and extend the model to new formulations and dose-escalation studies. Emerging efforts to compile PK parameters for combination therapies will facilitate this direction.

A second direction is deeper clinical contextualization. Our EHR and oncology cohort analyses show that combining \name with patient features improves adverse-event prediction over drug information alone, and richer data covering comorbidities, concomitant medications, and genomic profiles could improve it further. \revision{Future work will use longitudinal cohorts and time-to-event modeling to extend predictions to dynamic treatment courses.} Such data would admit variable dosing schedules, administration sequences, and drug holidays into the model, and these features describe how combinations are given in practice, where personalized dose adjustments have improved safety~\cite{Wu_2021_dosage}. Many drug classes appear in the training data through one or two compounds, which leaves the model without a drug class pattern to generalize from. Retrieval over preclinical literature within foundation models~\cite{queen2024procyon,su2024knowledge,gao2024empowering} could provide evidence for novel compounds from the same class at prediction time.

\name is a multimodal model for translational pharmacology of drug combinations. It predicts interactions between approved compounds and between approved and novel compounds.  \revision{\name scores a combination before any experimental test of that combination, which can guide the design of preclinical studies and clinical trials.}

\clearpage

\spacing{1}
\xhdr{Acknowledgements} 
We thank Jeffrey Tyner and Christopher Eide for their help with the BeatAML dataset and its data processing.
We thank Man Qing Liang and Xiang Zhang for their helpful discussions on the TWOSIDES dataset and Payal Chandak for discussions on the PrimeKG knowledge graph. We thank Philip Isola, Shanghua Gao, Huan He, Wenxian Shi, and Michelle M. Li for their feedback on model development. 
We thank Ayush Noori for discussions on using \name for personalized clinical outcome prediction. We appreciate Walker Rickord's comments on the manuscript. We thank Nigel Greene, Hebatallah Mohamed, and Michaël Ughetto for helpful feedback on \name and analyses. We thank Jane Knöchel for interpreting model predictions in MASH. We also thank Diansong Zhou and Karthick Vishwanathan for their valuable insights into the analysis of cancer drug combinations.
We gratefully acknowledge the support of NIH R01-HD108794, NSF CAREER 2339524, US DoD FA8702-15-D-0001, Harvard Data Science Initiative, Amazon Faculty Research, Google Research Scholar Program, AstraZeneca Research, Roche Alliance with Distinguished Scientists, Sanofi iDEA-iTECH, Pfizer Research, Gates Foundation (INV-079038), Chan Zuckerberg Initiative, John and Virginia Kaneb Fellowship at Harvard Medical School, Biswas Computational Biology Initiative in partnership with the Milken Institute, Harvard Medical School Dean's Innovation Fund for the Use of Artificial Intelligence, and Kempner Institute for the Study of Natural and Artificial Intelligence at Harvard University.  Any opinions, findings, conclusions or recommendations expressed in this material are those of the authors and do not necessarily reflect the views of the funders.

\xhdr{Data availability}
Data and results of our analyses are shared via the project website at \url{https://zitniklab.hms.harvard.edu/projects/Madrigal}. Datasets are available at Harvard Dataverse repository at \url{https://doi.org/10.7910/DVN/ZFTW3J}. The BeatAML dataset is not publicly available due to patient privacy. 
% \revision{
Clinical dataset from the Dana-Farber Cancer Institute is not publicly available due to patient privacy. The process of accessing the clinical dataset from Stanford Medicine is described at \url{https://stanford.redivis.com/datasets/53gc-8rhx41kgt}. The PDXE data can be accessed via \url{https://www.nature.com/articles/nm.3954}. 
% }

\xhdr{Code availability}
Python implementation of \name is available via the project website at \url{https://zitniklab.hms.harvard.edu/projects/Madrigal}. The code to reproduce results with examples of usage is at \url{https://github.com/mims-harvard/Madrigal}. 

\xhdr{Author contributions} 
Y.H. developed and implemented \name and designed evaluation setup. 
Y.H. retrieved and processed multimodal drug datasets used to train \name models and performed detailed analyses of \name's algorithm and model evaluation. 
V.U. implemented first-stage pretraining of \name for three modalities and integrated \name with a large language model. 
X.S., V.U., and Y.H. implemented alternative methods for benchmarking. 
N.H. retrieved and processed cell viability data. 
Y.H., X.S., and I.L. performed analyses on cancer combination therapies. I.L. processed datasets on T2D and cancer combination therapies, and Y.H. and I.L. performed analyses on metabolic disorders. 
I.M. retrieved and processed the independent oncology cohort data. X.S. retrieved and processed the longitudinal event-time cohort data. Y.H., X.S., and I.M. performed analyses to predict personalized clinical outcomes. 
L.C., D.O., B.J., R.J., M.G., and A.G.~provided expertise on clinical pharmacology and safety of drug combinations. L.C., D.O., and B.J.~provided additional expertise on the development of new machine learning methods, design of evaluation setup, benchmarking analyses, as well as real-world applications. 
M.Z.~designed and led the study. 
All authors contributed to writing the manuscript. 

\xhdr{Competing interests} 
L.C., D.O., and M.G. are employees and stockholders of AstraZeneca. B.J. performed this research while he was employed by AstraZeneca. The remaining authors declare no competing interests.

\clearpage
\pagestyle{empty}
% \section*{Figure Captions}

\clearpage

\clearpage

\clearpage

\clearpage

\clearpage 

\pagestyle{plain}
\customspacing{1}
\section*{Online Methods}

The Methods section describes (1) the data used in model development, benchmarking, validation, and case studies, (2) the model architecture and its implementation, (3) the benchmarking setup, and (4) applications of \name in translational pharmacology.

\section{Datasets}\label{method:data}

This section describes the drug combination outcomes \name is trained on, the four compound modalities it takes as input, the external datasets used to evaluate its predictions against organ-specific toxicity annotations and head-to-head clinical trials, and the patient-derived models and patient cohorts used in the pharmacological applications.

\subsection{Drug Combination Datasets}

We collect datasets from TWOSIDES (2019-11-13)~\cite{twosides} and DrugBank (2023-01-04)~\cite{drugbank}. 
TWOSIDES is a database derived from the FDA Adverse Event Reporting System (FAERS). FAERS is a comprehensive repository of adverse event and medication error reports submitted to the FDA. 
Following standard safety signal-detection criteria~\cite{gosho2017, noguchi2019}, we retain data with: (1) a minimum of three reports for the pair of drugs that report the side effect; (2) a proportional reporting ratio of at least 2; (3) a mean reporting frequency of 0.01 or higher; and (4) a Chi-square test statistic of 3.841 or higher ($p$-value $< 0.05$, with propensity-score matched drugs). 
Applying these criteria and filtering out safety outcomes with fewer than 500 samples~\cite{nyamabo2022}, we have compiled a total of 4,656,138 samples, which include 1,457 unique drugs and 795 unique safety outcomes. 

For DrugBank~\cite{drugbank}, we extract raw drug interaction statements from the XML dump and extract drugs and the safety outcome from those statements with manually specified regular expression patterns. The extracted safety outcomes are then manually examined, those that differ by rephrasing are grouped, and the directionality between the two drugs in the statement is neutralized. For example, ``@Drug1 increase the QTc-prolonging activities of @Drug2" is grouped with ``The risk or severity of QTc prolongation can be increased when @Drug1 is combined with @Drug2". We filter extracted data to (1) only contain small-molecule drugs with valid SMILES and (2) only include safety outcomes that have at least 20 drug pairs annotated.
Applying these criteria, we have compiled 1,188,371 samples, covering 3,632 drugs and 158 unique safety outcomes. 

\revision{
% Supplementary Fig.~S2 shows the number of drugs with different modality availability, the distributions of the number of drug pairs per unique safety outcome, and the number of outcomes per organ. 
Supplementary Fig.~\ref{fig:supp_fig_2a}{c,d} shows the distributions of the number of drug pairs per unique safety outcome and the number of outcomes per organ; modality availability across drugs is shown in Supplementary Fig.~\ref{fig:supp_fig_1}{c}.
}

\subsection{Drug Data Modalities}
We consider four data modalities:

\begin{enumerate}
    \item \textit{Structure} (struc.): As we focus on small-molecule compounds, structural information is essential and is universally available, represented by SMILES strings. These strings are converted into unique molecular graphs using the \texttt{RDKit} package~\cite{rdkit}.
    \item \textit{Pathways-based KG} (PKG): We incorporate biomedical knowledge at the pathway level from the drug-centric precision medicine knowledge graph, PrimeKG \cite{primekg}. This resource provides interaction profiles between approved drugs and diseases or proteins, and higher-order interactions with biological pathways. We exclude all drug-drug interactions and drug-phenotype interactions (individual drug side effects) to avoid information leakage.
    \item \textit{Cell viability profile upon drug perturbation} (CV): We utilize cell viability profiles from the PRISM Repurposing 19Q4 dataset \cite{prism} available at \href{https://depmap.org/portal/download/all/}{DepMap}. This dataset includes cell-line screens of chemical perturbation viability for 4,518 compounds against 578 cell lines. We use the preprocessing pipeline from \cite{Pan2022}, resulting in a 559-dimensional characteristic vector for each compound, with each entry corresponding to the change in viability of a cell line. 
    % \yepeng{Refer repreprocessingollowing data preprocessing file: \href{https://docs.google.com/document/d/1qkiVB35YVdrZ9SIXBtVyatWnnDd0cuZRmAIuGZt9bKc/edit}{Google doc}}
    \item \textit{Transcriptomics profile upon drug perturbation} (trans.): We gather transcriptomics profiles from the Extended CMap 2020 dataset available at \href{https://clue.io/data/CMap2020}{Connectivity Map} \cite{cmap}. We apply a quality control pipeline adapted from \cite{Zhu2021Tx}. We select 16 cell lines based on data availability (more than 500 compounds) and biological diversity (tissue types, cell lineage and primary disease), and treat each cell line as a separate modality. For each compound, within each cell line and treatment time, following recommendations in \cite{cmap_highest_dose}, we select the profiles from the maximal dosage applied. We adapt the pipeline in \cite{Zhu2021Tx}, where molecules are selected if they have more than two level-5 signatures in total (irrespective of the cell line, treatment time, dose, and plate). Repetitions and plates are averaged. We use profiles for each compound at the 24h treatment time, due to their high availability in all cell lines; the number of compounds in this group is usually the highest among all treatment time groups. The resulting input from each modality is a 978-dimensional feature vector, with each entry corresponding to the expression change of a landmark gene at one time stamp. 
\end{enumerate}

We match data from these modalities based on the \texttt{RDKit}-transformed canonical SMILES, each uniquely identifying a compound. Drug interaction data are mapped to compounds via DrugBank ID. Fig.~\ref{fig:figure_1}c and Supplementary Fig.~\ref{fig:supp_fig_1} provides an overview of the availability of drug data across these modalities.

\subsection{Drug-Induced Liver Injury (DILI) Dataset}\label{method:dili_dataset}

Using drug names, we match drugs in the DILI dataset curated by~\cite{Chen_2016_DILI} to DrugBank identifiers in our data. This yields 262, 234, 217, and 159 drugs with moderate, no, ambiguous, and most severe DILI concerns, respectively. 

\subsection{Drug-Induced Cardiotoxicity (DICT) Dataset}\label{method:dict_dataset}

Using drug names, we match drugs in the DICT dataset curated by~\cite{Qu_2023_DICT} to DrugBank identifiers in our data. This yields 301, 236, 68, and 206 drugs with moderate, no, ambiguous, and most severe DICT concerns, respectively. 

\subsection{Drug-Induced QTc Prolongation (DIQTA) Dataset}\label{method:diqta_dataset}

Using the DrugBank identifiers provided in the DIQTA dataset curated by~\cite{Li_2022_DIQTA}, we match its drugs to those in our data. This yields 55, 100, 241, and 109 drugs with moderate, no, ambiguous, and most severe DIQTA concerns, respectively. 

\subsection{Type 2 Diabetes, Heart Failure, and MASH Therapies}\label{method:comorbidity-dataset}

% We have also curated an extensive dataset of disease comorbidities used in our case studies, where we examine specific diseases, such as Type 2 Diabetes (T2D). The main comorbidities dataset is created by combining pre-existing datasets from FAERs \cite{faers_comorb} and Type 1 and Type 2 Diabetes datasets \cite{diabetes_comorb}. The FAERs dataset consists of common disease comorbidities extracted from FDA's Adverse Event Reporting System using Association Rule Mining (2014 - 2017), with 25,216 directed disease pairs (12,608 unique pairs) involving 1,059 unique diseases.  Type 1 and Type 2 diabetes comorbidities were extracted from Austria patients from 2006 - 2007; comorbidities were calculated through risk ratios (RR) in age strata from 20 to 65 years, and a disease was considered comorbid when its RR exceeded 2.0 in every age stratum. The T1D and T2D datasets cover 829 and 937 unique diseases (including the index disease), of which 391 and 265 are comorbid with T1D and T2D, respectively. 
% However, when examining specific diseases (MASH, heart failure, and kidney diseases), we note that they appear in our curated comorbidity dataset but are not included in PrimeKG \cite{primekg} due to compatible identification mapping. Thus, not all diseases in a specific class are examined, even though they might appear in the comorbidities dataset; only diseases in our comorbidity dataset and PrimeKG are used in the following analyses.

T2D medications are sourced from DrugCentral through PrimeKG~\cite{primekg}, the FDA Orange Book, and supplemented with mechanism of action data from UCSF~\cite{ucsf_t2d}, Mayo Clinic~\cite{mayo_t2d}, Cleveland Clinic~\cite{cleveland_t2d}, and DrugBank~\cite{drugbank} (Supplementary Table~\ref{tab:t2d_drugs}). We ensure that all medications, if not approved in the US, are marketed in Europe or Japan. 
% Although we focus primarily on small-molecule drugs and exclude insulin and its analogs, we include GLP-1 receptor agonists such as semaglutide, cotadutide, and lixisenatide, with available SMILES data.
Due to our focus on small molecule drugs, insulin and its analogs are excluded from this analysis.
% The complete list of drugs is shown in Supplementary Table~\ref{tab:t2d_drugs}. 
Similarly, HF medications are sourced from the American Heart Association and the FDA Orange book. The complete list of drugs is shown in Supplementary Table~\ref{tab:hf_drugs}.

We obtain the MASH clinical candidates and approved drugs, including monotherapies~\cite{Harrison_2023_clinical_landscape_nash,nash_approval} and combination therapies~\cite{Harrison_2023_nash_challenges,Suri_2022_nash_combos,Tilg_2023_nash_challenges}. The MOAs and clinical phases are manually annotated by extracting information from the references above. Only small-molecule drugs with valid SMILES are considered. The specific candidates included and their MoAs are shown in Supplementary Table~\ref{tab:mash_drugs}, and also in Fig.~\ref{fig:figure_5}f and Supplementary Fig.~\ref{fig:supp_fig_6}{b}.

\subsection{Drug Combination Clinical Trials Dataset}
\label{method:clinical_trials_data}
% \revision{
To compare predictions with head-to-head trial data, we systematically extract clinical trials concurrently investigating more than one combination. We extract AE data from clinical trials registered on \url{clinicaltrials.gov} through the Continuous Drug Combination Database (CDCDB)~\cite{Shtar_2022_CDCDB} (originally sourced via AACT; version April 16, 2024). Two trials not in CDCDB were additionally identified by a literature search assisted by OpenAI Deep Research (o3) and verified manually. To ensure valid comparisons of safety, we apply the following criteria for selecting appropriate trials: 
\begin{itemize}
    \item Is above Phase I;
    \item Started after year 2000;
    \item Has AE data;
    \item Has at least 20 participants per arm on average;
    \item Has at least two arms with exactly two different small molecule drugs that can be mapped to our DrugBank identifiers; each arm has a safety population size of at least 20.
\end{itemize}

We next compare AE data between each arm pair (pairs of arms with exactly two different small molecule drugs and a safety population size of at least 20). We apply three criteria to obtain the data for comparison for each trial (arm pairs and AEs):
\begin{itemize}
    \item The AE incidences significantly differ between arms with a two-sided Fisher's exact test after Bonferroni correction (adjusted $p$-value $<$ 0.05);
    \item At least one of the two arms had $\geq$3 affected participants with an incidence $\geq$1\%;
    \item The two arms should have comparable patient backgrounds and no design confounders such as crossover.
\end{itemize}
Applying these three criteria, we identify 35 trials (Supplementary Table~\ref{tab:trial_arms}). For trials that contain several arms administering the same drug combination (e.g., at different institutions), we manually confirm that their safety trends were concordant. We thus deem the comparison between that combination and any other combination arm significant for an AE if at least one of the same-combination arms meet the statistical threshold. 
% }

\subsection{BeatAML Ex Vivo Drug Synergy Dataset}\label{method:beataml-dataset}

We obtain the latest BeatAML ex vivo drug synergy dataset courtesy of Dr. Jeffrey Tyner, which is an updated dataset of similar outcome measurement as in~\cite{Bottomly_2022_BeatAML,Eide_2023_Venetoclax}, comprising more patients and drug combinations tested. The data preprocessing was done in the same approach as in~\cite{Bottomly_2022_BeatAML}, courtesy of Dr. Christopher Eide. We further filter the data so that only patient samples with RNA-seq profiles and small-molecule drug information are included. This gives us 336 patient samples, 135 drug combinations, and 12,161 (patient sample, drug combination) pairs.

Following the original BeatAML paper~\cite{Eide_2023_Venetoclax}, the synergy measure we use is combination ratio (CR), defined as the AUC (percentage of max) of the drug combination divided by the minimum AUC of each drug in the combination. A CR lower than 1 represents synergy pairs, and vice versa. The drugs in the dataset are matched with DrugBank ID based on their names.

\subsection{Patient-derived Xenograft Drug Combination Dataset}

We obtain the patient-derived xenograft encyclopedia (PDXE) dataset from~\cite{Gao_2015_pdx}. We further filter the data so that only patient samples with RNA-seq profiles and small molecule drugs with structural information available are included. This gives us 171 models, 11 drug combinations, and 366 (model, drug combination) pairs.

The efficacy measures we use are TimeToDouble, which corresponds to progression-free survival (time until tumor volume reaches 200\% of baseline), and BestAvgResponse, which corresponds to response (minimum value of the average of $\Delta {Vol}_t$ from t $=$ 0 to T, for T $\geq$ 10 d). The drugs in the dataset are mapped to DrugBank ID by name and through manual confirmation with literature.

\subsection{Longitudinal Event-Time Cohort}
\label{method:ehrshot_data}

% \revision{
We extract the cohort from EHRSHOT~\cite{ehrshot}, which contains de-identified structured data (e.g., diagnosis and procedure codes, medications, lab values) from EHRs of 6,739 patients from Stanford Medicine. EHRSHOT is longitudinal and includes data beyond ICU and emergency department patients.

For each labelled prediction time, we assemble the diagnoses, procedures, and medications recorded strictly before that time within the containing encounter. Each patient is represented by a single snapshot, taken at the last labelled prediction time, so that no patient contributes more than one example. We exclude patients with fewer than two visits and those whose snapshot carries no label. For the AE prediction tasks, we use the four-level severity labels (normal, mild, moderate, severe) provided in~\cite{ehrshot}. This yields cohorts of 4,464, 4,280, 4,379, 4,295, and 4,507 patients for thrombocytopenia, hyperkalemia, hypoglycemia, hyponatremia, and anemia, respectively, and 4,303 patients for each of the 30-day re-admission and all-cause mortality tasks. 
% }

\subsection{Oncology Cohort}
\label{method:dfci_data}

% \revision{
Analyses of patient-level data from the Dana-Farber Cancer Institute were conducted with approval from the Dana-Farber Institutional Review Board under protocols 19-033 and 19-025. Both protocols were granted waivers of authorization under the Health Insurance Portability and Accountability Act (HIPAA).

We curate an oncology cohort from the Dana-Farber Cancer Institute in which patients were on first-line regimens that contained exactly two small-molecule oncology drugs from June 2015 to March 2025. Regimens are retained only when at least ten patients meet these criteria and neither drug is indicated for hematological malignancies (as hematological malignancies can confound hematological AE measurements). Patients with missing treatment-related ICD codes or diagnosed with hematological malignancies are excluded. We consider the following AEs: hematotoxicity, neuropathy, thromboembolism, renal impairment, and fluid and electrolyte imbalance. Each AE is mapped to a set of ICD-10 codes determined by an oncology expert (Supplementary Table~\ref{tab:ae_icd_mapping}). These ICD-10 codes recorded within 28 days from the start of the first cycle of a regimen are flagged as regimen-related AEs. In total, we curate 3,577 patients with 26 unique regimens and 13 AE types (Supplementary Table~\ref{tab:ae_icd_mapping}). For each patient, in addition to regimen and ICD-based tumor tissue type, we also include age, gender, and palliative intent of treatment.
% }

\section{\name Model}
\label{method:framework}

Most compounds are observed in only a subset of data modalities, and many carry no combination safety labels. \name uses these partially characterized compounds during pretraining, aligning the modality-specific representations of each compound before finetuning, so that the observed modalities inform the missing ones. Alignment places the representations of one compound close together in the latent space regardless of which encoder produced them, so the representation computed from the observed modalities carries the information those modalities share with the missing ones, and the fusion module receives an input that reflects more than the observed data alone. 

\subsection{Problem Setup and Notation}
Let $\mathcal D = \{d_i\}_{i=1}^{n_{D}}$ denote the set of compounds with at least one modality or at least one combination outcome available. For the subset of drugs in $\mathcal D$ that have combination safety information available, a sample is defined by $(d_1, d_2, r)$ where $d_1, d_2 \in \mathcal D$ are two compounds and $r \in \mathcal R$ is a type of combination outcome. 
Each compound $d_i \in \mathcal D$ is uniquely identified by a SMILES string $x_i^{\mathsf{smiles}}$ and characterized by at most $n_M = 19$ modalities, namely:
\begin{itemize}
    \item struc.: represented by a molecular graph, $x_i^\mathsf{struc} = (\mathcal V_{x_i}, \mathcal E_{x_i}, \mathbf X_{x_i}, \mathbf E_{x_i})$ (generated from $x_i^{\mathsf{smiles}}$)
    \item PKG: represented by a drug node and its neighborhood or computation tree on a drug-centered knowledge graph $G$, $x_i^{\mathsf{PKG}} = (d_i, G)$
    \item CV: represented by a perturbation profile, $x_i^{\mathsf{CV}} \in \mathbb R^{559}$ 
    \item trans.-\{cell line\} (\{cell line\} denotes one of the 16 cell lines we collected, for example, MCF7): represented by a perturbation profile, $x_i^{\mathsf{trans.}\text{-}\mathsf{\{cell\ line\}}} \in \mathbb R^{978}$
    % \mathbf{x}_i^{\mathsf{trans.}\text{-}\mathsf{\{cell\ line\}}}
\end{itemize}

Denote the full set of modalities as $\mathcal M =$ \{$\mathsf{struc.}$, $\mathsf{PKG}$, $\mathsf{CV}$, $\mathsf{trans.}\text{-}\mathsf{MCF7}$, $\mathsf{trans.}\text{-}\mathsf{VCAP}$, $\mathsf{trans.}\text{-}\mathsf{PC3}$, $\mathsf{trans.}\text{-}\mathsf{A549}$, $\mathsf{trans.}\text{-}\mathsf{A375}$, $\mathsf{trans.}\text{-}\mathsf{HA1E}$, $\mathsf{trans.}\text{-}\mathsf{HT29}$, $\mathsf{trans.}\text{-}\mathsf{HCC515}$, $\mathsf{trans.}\text{-}\mathsf{NPC}$, $\mathsf{trans.}\text{-}\mathsf{HELA}$, $\mathsf{trans.}\text{-}\mathsf{HEC108}$, $\mathsf{trans.}\text{-}\mathsf{THP1}$, $\mathsf{trans.}\text{-}\mathsf{HEPG2}$, $\mathsf{trans.}\text{-}\mathsf{YAPC}$, $\mathsf{trans.}\text{-}\mathsf{ASC}$, $\mathsf{trans.}\text{-}\mathsf{HUVEC}$\}. Also, denote the set of modalities available to a compound $d$ as $\mathcal M_d \subseteq \mathcal M$. For each modality $m \in \mathcal{M}$, a modality-specific encoder $f^m: \mathcal X^m \to \mathbb R^{\mathrm{shared}} $ maps the modality-specific data to representations in a shared latent space. Note, for simplicity, through the following development and in the main text, we denote transcriptomics as one modality ($\mathsf{trans.}$) to enhance readability and clarity, while it is treated as 16 modalities (each cell line as one) in the model (also see Supplementary Note~\ref{note:architecture}; Supplementary Fig.~\ref{fig:supp_fig_1}c). 

\subsection{Model Architecture}
Our model architecture, designed to handle any composition of compound modalities as input and predict safety profiles for drug combinations, is depicted in Fig.~\ref{fig:figure_1}. The model's encoder components are first initialized and adapted with encoder-specific pretext tasks. They are then pretrained with a contrastive objective and transferred to the downstream finetuning for combination safety prediction. The model architecture and learning objectives are formally defined in the following subsections and optimized sequentially in three stages. 

\subsubsection{Modality-Specific Encoders}
\label{method:encoder_init}

Each modality is encoded by an architecture matched to its data type. A Heterogeneous Graph Transformer~\cite{hgt} encodes the pathways knowledge graph, a Graph Isomorphism Network~\cite{xu2019} encodes the molecular structure, and a multilayer perceptron encodes the cell viability profile, of which we retain only the encoder at inference. A  chemCPA~\cite{hetzel2022} encoder, configured with RDKit descriptors and without dosage, encodes the transcriptomic profiles of all 16 cell lines. We initialize every encoder $f^m$, $m\in \mathcal M$, from scratch and adapt it with a modality-specific pretext task, so that each encoder produces informative representations before alignment. Aligned initialization also avoids modality competition, in which encoders initialized at random converge to a solution dominated by one modality while the remaining modalities contribute little~\cite{Huang2022}.
\begin{itemize}
    \item struc.: A supervised property prediction task is used to train the structure encoder. Let $y^{\mathsf{struc.}} \in \{0, 1\}$ denote the binary activity label of compound $d$ for a property of interest. We apply a linear head $h^{\mathsf{struc.}}$ above the structural encodings to predict 17 properties for about 90k molecules from PubChem BioAssay \cite{wang2012pubchem} compiled by the MoleculeNet benchmark as the Maximum Unbiased Validation (MUV) dataset \cite{wu2018moleculenet}. We then optimize $f^{\mathsf{struc.}}$ by minimizing a binary cross-entropy loss over the $n_{\mathsf{struc.}}$ labeled (compound, property) pairs, i.e.
    \begin{align*}
    L_{\mathsf{BCE}}^{\mathsf{struc.}} = -\frac{1}{n_{\mathsf{struc.}}} \sum_{i=1}^{n_{\mathsf{struc.}}} \Big( & y_i^{\mathsf{struc.}} \log \sigma\big(h^{\mathsf{struc.}}(f^{\mathsf{struc.}}(x_i^{\mathsf{struc.}}))\big) \\
    & + (1 - y_i^{\mathsf{struc.}}) \log \big(1 - \sigma(h^{\mathsf{struc.}}(f^{\mathsf{struc.}}(x_i^{\mathsf{struc.}})))\big) \Big),
    \end{align*}
    where $\sigma$ is the sigmoid function.
    \item PKG: A self-supervised knowledge graph link prediction task is used to train the pathways encoder. In this task, we predict the existence of edges between two nodes in the knowledge graph $G$ (removing all drug-drug and drug-phenotype edges). Let $E$ denote all such edges, $N$ denote negative samples and $h^{\mathsf{PKG}}$ denote the scoring function for link prediction from a triplet of $(f^{\mathsf{PKG}}(s, G), f^{\mathsf{PKG}}(t, G), r)$, where $s$ is the source node, $t$ is the target node and $r$ is the edge type. We optimize $f^{\mathsf{PKG}}$ for minimizing a binary cross-entropy loss, i.e.
    \begin{align*}
    L_{\mathsf{BCE}}^{\mathsf{PKG}} = \frac{1}{n_{\mathsf{PKG}}} \Big( & - \sum_{(s, t, r) \in E} \log h^{\mathsf{PKG}}(f^{\mathsf{PKG}}(s, G), f^{\mathsf{PKG}}(t, G), r) \\ 
    & - \sum_{(s, t, r) \in N}
    \big(1 - \log h^{\mathsf{PKG}}(f^{\mathsf{PKG}}(s, G), f^{\mathsf{PKG}}(t, G), r)\big)\Big)
    \end{align*}
    \item CV: A reconstruction~\cite{bank2020autoencoders} objective is used to train the CV encoder. The encoder compresses the input information into a latent representation from which the decoder reconstructs the input. 
    Specifically, $f^{\mathsf{CV}}$ encodes $x^{\mathsf{CV}}$ as latent vector $z^{\mathsf{CV}}$, while $h^{\mathsf{CV}}$ decodes $z^{\mathsf{CV}}$ to reconstruct $x^{\mathsf{CV}}$.
    In practice, we train the model to minimize the mean square error loss: 
    $$
    L_\mathsf{MSE}^{\mathsf{CV}} = \frac{1}{n_{\mathsf{CV}}} \sum_{i=1}^{n_{\mathsf{CV}}} 
    (h^{\mathsf{CV}}(f^{\mathsf{CV}}(x_i^{\mathsf{CV}}))-x_i^{\mathsf{CV}})^2
    $$
    \item trans. (trans.-\{cell lines\}): We pretrain the encoder with a strategy similar to chemCPA but omit the adversarial drug loss, since we learn drug representations rather than counterfactual predictions. We refer interested authors to \cite{hetzel2022} for details about their training objective.
\end{itemize}

\subsubsection{Multimodal Alignment through Contrastive Learning}

The objective is to align the representations generated for the same drug from different modalities with the structure modality with a multimodal contrastive representation learning objective. We adopt the InfoNCE objective~\cite{oord_2018_cpc} with minor modifications similar as in~\cite{Chen2020}, and jointly learn all encoders $f^m$ for all modalities $m$ using the following loss:
$$
L_{\mathsf {cont}} = \sum_{m \neq \mathsf{struc.}} L\left(m, \mathsf{struc.}\right) = \sum_{m \neq \mathsf{struc.}} \left(\ell\left(m, \mathsf{struc.}\right) + \ell\left(\mathsf{struc.}, m\right)\right),
$$
where
$$
\ell\left(m_u, m_v\right)=-\sum_{i=1}^B \log \frac{\mathrm{sim}_{(m_u, m_v)}(d_i, d_i)}{\sum_{j=1}^B\left(\mathrm{sim}_{(m_u, m_v)}(d_i, d_j) + \mathbf{1}_{j \neq i}\cdot\mathrm{sim}_{(m_u, m_u)}(d_i, d_j) \right)},
$$
% $$
% \ell\left(m, \mathsf{struc.}\right)=-\sum_{i=1}^B \log \frac{\mathrm{sim}_{(m, \mathsf{struc.})}(d_i, d_i)}{\sum_{j=1}^B\left(\mathrm{sim}_{(m, \mathsf{struc.})}(d_i, d_j) + \mathbf{1}_{j \neq i}\cdot\mathrm{sim}_{(m, m)}(d_i, d_j) \right)},
% $$
and
$$
\mathrm{sim}_{(m_u, m_v)}(d_i, d_j) = \mathrm{exp}(f^{m_u}(d_i) \cdot f^{m_v}(d_j) / \tau),
$$
is minimized.

We randomly sample the other modality (other than structure) for each compound, with the probability inversely proportional to the modality's availability, measured by its prevalence in the pre-training set of compounds, preventing well-covered modalities from dominating the objective.
\revision{Modality availability is recorded per compound as a binary mask, and sampling is restricted to the modalities a compound has. Only compounds with at least two available modalities enter this stage. The structure modality is available for every compound, and the other modality is drawn from that compound's remaining available modalities.}

\revision{
The alignment reconciles heterogeneity across modalities. 
% The four data types differ in dimensionality, scale, and noise structure. A modality-specific encoder matched to each data type first maps it to a representation and is adapted with its own pretext task (stage 1), so the encoders absorb the differences in the raw data. Alignment then places these representations in a common latent space, the unit hypersphere, and compares them only through inner products of $\ell_2$-normalized embeddings with a shared temperature $\tau$, which removes differences in scale and geometry before any cross-modal operation. 
We anchor on structure instead of aligning all modality pairs. Structure is the only modality observed for every compound, so anchoring on it lets an observed modality inform a missing one for any compound, and it imposes agreement between the other modalities indirectly. Minimizing the sum of InfoNCE terms maximizes a lower bound on the mutual information between the structure representation and each other representation~\cite{poole_2019_variation_bounds_mi}, so alignment increases the information the modalities share. Alignment improves each modality-specific encoder used on its own (Supplementary Table~\ref{tab:before_after_CL}), and modalities still contribute beyond structure after alignment (Fig.~\ref{fig:figure_2}c,g; Supplementary Fig.~\ref{fig:supp_fig_2b}{a}), indicating that information transfers across modalities without modality collapse.
}

\subsubsection{Modality Fusion}
\label{method:modality-fusion}

We fuse modality representations with a Transformer encoder that uses attention bottlenecks~\cite{Nagrani_2021_bottleneck}. Attention-based fusion accepts a variable set of input tokens and performs well in other multimodal settings, including vision-language models~\cite{tsai2019, li_2021_align_before_fuse, Jaegle_2022_perceiver_io}. The 19 modality tokens are imbalanced, since transcriptomics contributes 16 of them, one per cell line. We therefore insert bottleneck tokens and restrict attention so that cell-line tokens attend among themselves and to the bottleneck tokens, and the remaining modality tokens do the same (Supplementary Note~\ref{note:architecture}). \revision{Routing all exchange between the two token groups through a small set of bottleneck tokens forces each group to compress the information it passes to the other, which prevents the 16 transcriptomic tokens from dominating fusion.
A learnable summarization query then cross-attends to the output tokens of the last fusion layer and produces the multimodal drug embedding~\cite{Recasens_2023_zorro, Jaegle_2022_perceiver_io}. The query attends to every token available for the drug, namely the observed non-transcriptomic modality tokens, the bottleneck tokens, and the observed transcriptomics cell-line tokens. 
We ablate the two design choices, the restriction of attention to the bottleneck tokens and the scope of the readout, across the two datasets and all four splits (Supplementary Table~\ref{tab:fusion_readout_ablation}), and we compare bottleneck-mediated fusion against unrestricted self-attention and against a learned gated weighted sum of the modality embeddings (Supplementary Table~\ref{tab:fusion_mechanism}).
A bilinear decoder $h^{\mathrm{dec}}$ scores a pair of compounds against each safety outcome, which keeps the prediction module inexpensive over 953 outcomes (Supplementary Note~\ref{note:architecture}), except for an experiment where we replace this decoder with LLM (Supplementary Note~\ref{note:llm_extension}).}

\subsubsection{Prediction Module}
\label{method:model_finetuning}

The fusion module $g^{\mathrm{fusion}}$ maps the modality-specific encodings available for compound $d$, $\{f^m(x^m)\}_{m\in \mathcal M_{d}}$, to a single compound embedding $\mathbf z^{\mathcal M_{d}} \in \mathbb R^{\mathrm{joint}}$. The module accepts any non-empty subset of the modalities, so a compound characterized by structure alone and a compound characterized by all four data types pass through the same computation and yield embeddings in the same space. Decoder $h^{\mathrm{dec}}: \mathbb R^{\mathrm{joint}} \times \mathbb R^{\mathrm{joint}} \times \mathcal R\to [0,1]$ then takes the embeddings of two compounds together with an outcome $r \in \mathcal R$ and returns the predicted probability that the pair produces that outcome. Let $S$ denote all samples and $S_{\mathrm{neg}}$ denote all negative samples. We jointly optimize the encoders $f^m$ together with $g^{\mathrm{fusion}}$ and $h^{\mathrm{dec}}$, initialized at random, by minimizing the binary cross-entropy loss:
\begin{align*}
L_{\mathsf{BCE}}^{\mathrm{ft}} = \frac{1}{|S|} \Bigg( & - \!\!\!\!\!\sum_{(d_1, d_2, r) \in S} \log h^{\mathrm{dec}} \big(g^{\mathrm{fusion}}(\{f^m(d_1^m)\}_{m\in \mathcal M_{d_1}}), g^{\mathrm{fusion}}(\{f^m(d_2^m)\}_{m\in \mathcal M_{d_2}}), r \big) \\ 
& - \!\!\!\!\!\!\!\sum_{(d_1, d_2, r) \in S_{\mathrm{neg}}} \!\!\!\!\!\!\!\Big(1 - \log h^{\mathrm{dec}} \big(g^{\mathrm{fusion}}(\{f^m(d_1^m)\}_{m\in \mathcal M_{d_1}}), g^{\mathrm{fusion}}(\{f^m(d_2^m)\}_{m\in \mathcal M_{d_2}}), r \big) \Big) \Bigg)
\end{align*}
\revision{During finetuning, we enumerate the non-empty subsets of all modalities observed for each compound and draw one uniformly at random in every epoch, so that a compound is presented under varying combinations of its own available modalities. We train each drug pair under two further configurations, one in which the structure modality alone represents each compound, and one in which the structure modality alone represents the first compound and the sampled subset represents the second. These configurations expose the model during training to the range of modality combinations it encounters at inference. Compounds with no modality beyond structure are always presented with structure alone. We never impute a missing modality, and the fusion module excludes its token from self-attention and from the summarization query.}

\subsection{Model Training}

We finetune encoders, fusion module, and prediction module, with encoders initialized from a pretrained model checkpoint. We use AdamW with linear warm-up and cosine annealing for all three stages and keep the model checkpoint with the highest validation AUPRC. 

\subsubsection{Hyperparameter Tuning}

We use Weights and Biases~\cite{wandb} to select hyperparameters via a random search over the hyperparameter space. The best-performing hyperparameters are selected by optimizing the AUPRC on the validation set. The hyperparameter space on which we perform a random search to choose the optimal set of hyperparameters is: 
position embedding $\in [\text{learnable}, \text{sinusoidal}]$, position embedding dropout $\in [0.1, 0.2, 0.4]$, number of heads in transformer encoder $\in [2, 4, 8]$, dimension of heads in transformer encoder $\in [64, 128, 256]$, number of layers in transformer encoder $\in [2, 3, 4, 6]$, dimension of feed forward layer in transformer encoder $\in [256, 512, 1024]$, dropout in transformer encoder $\in [0.1, 0.15, 0.2, 0.25, 0.3, 0.35, 0.4, 0.45, 0.5]$, number of attention bottlenecks $\in [2, 4]$, dropout in projector $\in [0.1, 0.2, 0.4]$, warmup epochs $\in [10, 20, 50, 100]$, learning rate (for each of structure encoder, pathways encoder, cell viability and transcriptomics encoder, fusion module, prediction module) $\in [1e$-$4, 2e$-$4, 5e$-$4, 1e$-$3, 2e$-$3, 5e$-$3]$, weight decay $\in [0.001, 0.01, 0.1]$, epsilon $\in [1e$-$8, 1e$-$7, 1e$-$6]$, whether or not having a separate adaptor when the drug only has one (structure) modality available, and the ordering of dropout and normalization layer (i.e.,~normalization before dropout, or vice versa).
    
To reduce cost, we only tune hyperparameters for each dataset using one splitting strategy (split-by-drugs (random)). The optimal sets of hyperparameters selected are:
\begin{itemize}
    \item DrugBank dataset: position embedding $= \texttt{sinusoidal}$, position embedding dropout $= 0.2$, number of heads in transformer encoder $= 8$, dimension of heads in transformer encoder $= 64$, number of layers in transformer encoder $= 3$, dimension of feed-forward layer in transformer encoder $= 256$, dropout in transformer encoder $= 0.15$, number of attention bottlenecks $= 4$, dropout in projector $= 0.1$, warmup epochs $= 100$, learning rate (for each of structure encoder, pathways encoder, cell viability and transcriptomics encoder, fusion module, prediction module) $= 1e$-$4, 1e$-$3, 1e$-$4, 1e$-$3, 1e$-$3$, weight decay $= 0.001$, epsilon $= 1e$-$6$, no separate projector for when the drug has only one (structure) modality available, and normalization layer first.
    \item TWOSIDES dataset: position embedding $= \texttt{learnable}$, position embedding dropout $= 0.1$, number of heads in transformer encoder $= 2$, dimension of heads in transformer encoder $= 256$, number of layers in transformer encoder $= 3$, dimension of feed-forward layer in transformer encoder $= 512$, dropout in transformer encoder $= 0.4$, number of attention bottlenecks $= 2$, dropout in projector $= 0.1$, warmup epochs $= 10$, learning rate (for each of structure encoder, pathways encoder, cell viability and transcriptomics encoder, fusion module, prediction module) $= 1e$-$3, 5e$-$4, 1e$-$3, 2e$-$4, 5e$-$3$, weight decay $= 0.001$, epsilon $= 1e$-$6$, separate projector for when the drug has only one (structure) modality available, and normalization layer first.
\end{itemize}

\subsubsection{Implementation}
We implement all \name models using PyTorch (Version~1.13.1)~\cite{pytorch}. \name models are trained on a single NVIDIA A100 GPU. For the patient-level analyses, the BeatAML synergy predictor (a symmetric bilinear decoder followed by an MLP) is implemented in PyTorch, while the random forest models used for PDXE and for the independent oncology cohort are implemented using scikit-learn~\cite{pedregosa2011scikit}; all are trained on the CPU.

\name is computationally efficient due to the relatively small computation overhead and the feasibility of fitting the entire dataset to a single GPU with efficient operations, taking about one hour to fine-tune on the DrugBank data set on one GPU, compared to the longer runtime required by a few strong models (Supplementary Table~\ref{tab:runtime}).

\section{Benchmarking \name Model}

\subsection{Dataset Splits}
\label{method:benchmark_dataset_split}
As novel compounds typically lack combination safety information, we held out (about) 20\% approved drugs in each dataset, along with their associated combinations, to rigorously evaluate model performance under realistic conditions. To achieve this, we designed three distinct testing scenarios using different data splitting strategies: based on drug ATC codes (``split-by-drugs (ATC)"), drug targets (``split-by-drugs (targets)"), random splits by drugs (``split-by-drugs (random)"). In each splitting setting, we also additionally split (about) 10\% drugs into a validation set to prevent model overfitting and for hyperparameter tuning. 
In addition, to evaluate the model also in a transductive setting, in the fourth strategy, we randomly split 20\% drug pairs and all associated combinations into a test set (``split-by-drug pairs"), 10\% drug pairs and all associated combinations into validation set, and rest in the train set.

\revision{We distinguish drugs from samples throughout. Train, validation, and test drugs are the drugs assigned to each set; train, validation, and test samples are the (drug 1, drug 2, outcome) triplets assigned to each set. In the split-by-drug pairs setting, samples are assigned directly, and a drug may appear in both train and test samples. In the three split-by-drugs settings, the assignment of drugs determines the assignment of samples, as follows.}
In the three split-by-drugs settings, train samples are formed by selecting those samples where both drugs are in the train set. However, there is a unique aspect for validation or test samples in these settings: for each validation or test drug, the other drugs it interacts with could either be in the validation or test or included in the train set. These two types of samples have different implications: one scenario mimics the case where both drugs are novel compounds, while the other can be viewed as the scenario where one drug is a novel compound and the other is an approved drug. In practice, it is more valuable to understand the interaction profiles of a novel compound with approved drugs in related therapeutic areas (such as comorbidities). Therefore, we focus on the latter group of samples when evaluating the model. Specifically, validation samples are formed by selecting those samples where one drug is in the validation set while the other is in the train set; test samples are formed by selecting those samples where one drug is in the test set while the other is in the train or validation set. 

In the split-by-drugs (ATC) setting, drugs in each dataset are grouped according to the initial letter of their ATC codes (anatomical or pharmacological groups). ATC codes are randomly split into train, validation, and test sets. For the DrugBank dataset, drugs whose ATC codes start with ``N", ``V", ``J", ``B", ``C", ``A" are split into train set, containing a total of 2,562 drugs and 584,891 samples, drugs whose ATC codes start with ``D", ``L" are split into validation set, containing a total of 427 drugs and 162,608 samples, and drugs whose ATC codes start with ``G", ``H", ``M", ``R", ``P", ``S" are split into test set, containing a total of 613 drugs and 368,646 samples. For the TWOSIDES dataset, drugs whose ATC codes start with ``H", ``L", ``G", ``S", ``D", ``A", ``N", ``J", ``M", are split into train sets containing a total of 1,021 drugs and 2,084,566 samples, drugs whose ATC codes start with ``R", ``P" are split into validation set, containing a total of 148 drugs and 747,959 samples, and drugs whose ATC codes start with ``B", ``V", ``C" are split into test set, containing a total of 265 drugs and 1,478,489 samples.

In the split-by-drugs (targets) setting, because each drug can have multiple targets, which makes naively splitting targets infeasible, we construct a drug network where two drugs are connected if they share any target. The largest connected component (LCC) contains more than half (DrugBank dataset: 52\%; TWOSIDES dataset: 66\%) of drugs with target profiles, which we then detect communities (DrugBank dataset: 14; TWOSIDES dataset: 9) via the Louvain algorithm. The communities from LCC, along with other components in the network, are randomly split such that 20\% of drugs are in the test set, 10\% of drugs are in the validation set, and others are in the train set. All drugs without target information in DrugBank are split into train sets. Specifically, for the DrugBank dataset, 2,482, 423, 727 drugs are split into train, validation, and test sets, containing a total of 426,890, 264,272, 381,841 samples, respectively; for the TWOSIDES dataset, 987, 156, 314 drugs are split into train, validation, and test sets, containing a total of 1,923,741, 669,568, 1,706,453 samples, respectively.

In the split-by-drugs (random) setting, we randomly divided the drugs in each dataset into train, validation, and test sets with the ratios above. For the DrugBank dataset, the train, validation, and test sets contain 565,166, 170,888, and 388,492 samples, respectively; for the TWOSIDES dataset, the train, validation, and test sets contain 2,345,947, 664,265, and 1,432,496 samples, respectively.

In the split-by-drug pairs setting, we randomly divided drug pairs in each dataset into train, validation, and test sets with the ratios above. For the DrugBank dataset, the train, validation, and test sets contain 831,859, 118,837, and 237,675 samples, respectively; for the TWOSIDES dataset, the train, validation, and test sets contain 3,254,433, 466,311, and 935,394 samples, respectively.

\subsection{Negative Sampling and Class Imbalance}
\label{method:negatives}
\revision{
Following benchmarking setups in previous literature~\cite{Zitnik2018,nyamabo2022,Su_2022_ddkg,Su_2024_tiger}, for each positive (drug 1, drug 2, outcome) triplet, we generate two negative samples by replacing one drug with a randomly drawn drug, keeping the outcome fixed. Replacement drugs are drawn uniformly, and we exclude self-pairs or false negatives. 

For train samples, we replace each of the two drugs in turn, giving (drug 1, drug 2', outcome) and (drug 1', drug 2, outcome). The same applies throughout the split-by-drug pairs setting to those validation and test samples. For the validation and test samples we report in the split-by-drugs settings, which contain exactly one validation or test drug (Section~\ref{method:benchmark_dataset_split}), we instead keep that drug in place and draw two independent replacements for the other drug, giving (drug 1, drug 2', outcome) and (drug 1, drug 2'', outcome). That drug is therefore present in the positive sample and in both of its negatives, so every evaluated sample remains a prediction for a validation or test drug. For the test samples we report, replacement drugs are drawn from the train and validation drugs, so a test drug is never introduced into a negative sample. Negatives are generated once under a fixed seed, stored with the splits, and shared across all methods and all seeds, giving a positive-to-negative ratio of exactly 1:2.

Because each negative inherits the outcome label of the positive it is derived from, the ratio holds within every outcome individually: each outcome has a positive prevalence of exactly $1/3$, in both datasets and in every split. Class balance is thus fixed by construction and identical across outcomes, and we consequently train with an unweighted binary cross-entropy loss (Section~\ref{method:model_finetuning}). Imbalance across outcomes, where some outcomes are far more frequent than others, is handled instead by macro-averaging every metric within each outcome before averaging over outcomes (Section~\ref{method:eval_metrics}), so that each outcome contributes equally regardless of its support. We additionally report AUPRC and Fmax alongside AUROC, both of which are sensitive to performance on the positive class.
}

In certain analyses, we also calculate metrics for each test drug. This is done by calculating metrics within all test samples containing the test drug and all corresponding negative samples.

\subsection{Experimental Setup}

\revision{
\name is trained with five different seeds (0, 1, 2, 42, 99) for the split-by-drugs (targets), split-by-drugs (ATC), and split-by-drug pairs splitting strategies and with three different seeds (0, 42, 99) across three independent repeats of random splits (in total, 9 runs) for the split-by-drugs (random) splitting strategy. Performance is averaged with standard deviations presented in all benchmarking tables. 
To mimic realistic data availability for novel compounds, in the three split-by-drugs settings, at evaluation time, available modalities are masked such that only structure and transcriptomic modalities are available to the validation and test drugs, while the train drug it is paired with keeps all of its available modalities. In the split-by-drug pairs setting, all available modalities are used for both drugs. This applies to \name and its ablations only; existing methods use their native inputs.

We also ablate \name in three ways:
\begin{enumerate}
    \item W/o CL: Finetuning on the combination safety dataset without the modality alignment stage.
    \item Struc. only: Only using structure modality during model finetuning (but with all modalities during modality alignment).
    \item Struc. only w/o CL: Finetuning on the combination safety dataset with only the structure modality and without the modality alignment stage.
\end{enumerate}
Each modality-specific encoder is initialized and adapted with its own pretext task (Methods Sec.~\ref{method:encoder_init}) before model finetuning.
}

\subsection{Performance Metrics}
\label{method:eval_metrics}
We evaluate model performance with standard classification metrics, including the area under receiver-operating curve (AUROC), the area under the precision-recall curve (AUPRC), and maximum F-measure (Fmax), calculated in a ``macro" manner, i.e., within each label (safety outcome) then averaged. We macro-average because micro-averaging pools predictions across outcomes with different base rates. 

Specifically, for each outcome, given predicted scores $\mathbf{s} = (s_1, s_2, \dots, s_N)$ and corresponding binary labels $\mathbf{y} = (y_1, y_2, \dots, y_N)$, Fmax is calculated as:
$$
F_{\max} = \max_{\tau} F(\tau),
$$
where the F1 score at threshold $\tau$ is defined as
$$
F(\tau) = \frac{2\, \text{Prec}(\tau) \, \text{Rec}(\tau)}{\text{Prec}(\tau) + \text{Rec}(\tau)},
$$
with the precision and recall at threshold $\tau$ being
$$
\quad
\text{Prec}(\tau) = \frac{\sum_{i=1}^N y_i\, \mathbbm{1}(s_i \geq \tau)}{\sum_{i=1}^N \mathbbm{1}(s_i \geq \tau)},
\quad
\text{Rec}(\tau) = \frac{\sum_{i=1}^N y_i\, \mathbbm{1}(s_i \geq \tau)}{\sum_{i=1}^N y_i},
% \quad
% \mathbb{I}(s_i \geq \tau) =
% \begin{cases}
% 1, & \text{if } s_i \geq \tau, \\
% 0, & \text{otherwise.}
% \end{cases}
$$
where $\mathbbm{1}(\cdot)$ is the indicator function.

\subsection{Existing methods for drug-drug interaction prediction}
\label{method:benchmark_baselines}

\revision{
We compare \name with eight existing methods across two modalities on two datasets. These models use either late fusion~\cite{Su_2022_ddkg,muffin,Su_2024_tiger,Jiang2024deepgcl,Li2026llmddi}, where each drug molecule is encoded separately and merged, or early fusion, where molecular interactions are modeled from the start~\cite{Kim_2023_deepddi_paxlovid,caster,nyamabo2022}.

DeepDDI \cite{Kim_2023_deepddi_paxlovid} is a structure-based DDI prediction model. It uses a deep neural network to predict drug combinations from drug structural information. Following the authors, each drug is represented by the 50 leading principal components of its structural-similarity profile against the training drugs. We tuned the predictor on validation performance, arriving at a four-layer feed-forward network of hidden width 512, trained for 300 epochs with batch size 4000 and learning rate $10^{-4}$.

CASTER \cite{caster} is inspired by drug chemical substructures. It first extracts frequent substructures from a molecular database. Then, it designs a latent feature embedding module to represent drugs in terms of the extracted frequent substructures and predict drug combinations. We use the reference implementation in the ChemicalX library \cite{chemicalx}, in which each drug pair is encoded by the element-wise maximum of the two drugs' 1722-bit Morgan fingerprints, and we retain the library's default layer widths and dictionary-learning loss coefficients. We train for 40 epochs and select the checkpoint with the best validation AUPRC, using batch size 256 and Adam learning rate $10^{-3}$ as specified in the original paper. 

GMPNN-CS \cite{nyamabo2022} predicts drug combinations by learning chemical substructures with different sizes and shapes from the molecular graph representations of drugs. It considers the edge between atoms as gates that control the flow of message passing and, therefore, delimit the substructures in a learnable way. We use the authors' implementation with their recommended settings of 10 message-passing iterations and hidden dimension 64, with dropout 0.2, trained for 200 epochs with batch size 1600 and learning rate $10^{-3}$.

DDKG \cite{Su_2022_ddkg} is a knowledge-graph-based DDI prediction model with attention-based neighborhood aggregation. Rather than initializing drug nodes at random, it derives their initial embeddings from SMILES strings through an encoder-decoder layer, while other knowledge-graph entities use learned embeddings, and then recursively propagates and aggregates information from neighbouring entities to obtain global drug representations. Neighbours are weighted by an attention mechanism conditioned on both the drug embedding and the relation type, so that the contribution of each neighbour is learned rather than fixed. Interactions are predicted from the resulting pair of drug representations. 

MUFFIN \cite{muffin} is a multi-scale feature fusion model that combines the internal chemical structure of a drug with its external semantic context in a biomedical KG. Molecular structure features are extracted from SMILES with a message-passing neural network, semantic features are obtained from the KG with TransE, and the two views are then crossed to capture both local and global signal, with fine-grained interaction features formed by an element-wise product before the concatenated representation is passed to the classifier. 

TIGER \cite{Su_2024_tiger} is a dual-channel, relation-aware heterogeneous graph transformer. One channel encodes the drug molecular graph and the other the biomedical knowledge graph, and a relation-aware self-attention mechanism captures the diverse semantic relations that hold between pairs of nodes in a heterogeneous graph. Embeddings obtained at both the graph level and the node level are combined, so that predictions draw on the structural properties of the drugs as well as the contextual information carried by the KG. Subgraph sampling is used to keep the method tractable on large knowledge graphs.

DeepGCL \cite{Jiang2024deepgcl} is a graph contrastive learning model for DDI prediction. It encodes each drug both from its molecular graph and from the drug's neighborhood in the DDI interaction network, and contrasts the two views during training, using their fused representation for prediction. We adapted the binary prediction head in DeepGCL to the multilabel outcome setting used in our benchmark. We compared training with a cross-entropy objective over outcomes against a binary cross-entropy objective matched to the outcome, and selected the former based on validation performance. The paper's batch size of 24 and learning rate of $10^{-3}$ diverge, so we tuned and selected batch size 1024 and learning rate $10^{-4}$ based on validation performance. The contrastive term is weighted by 0.1, same as the paper.

LLM-DDI \cite{Li2026llmddi} initializes each entity of a biomedical knowledge graph with a text embedding from a large language model, propagates these features with a relation-aware attention network, and scores a drug pair from the two final representations. We reimplemented the model from the published formula with the authors' reported settings: 2 propagation layers, 4 uniformly sampled neighbors per entity, and entity and relation embedding dimension 1536. We use the same PrimeKG as used in \name as the knowledge graph. Drug features are \texttt{gpt-5} descriptions generated based on DrugBank chemical and bioactivity fields (molecule type, SMILES, chemical classification, mechanism of action, and target, enzyme, carrier and transporter proteins), embedded with \texttt{text-embedding-3-large} at the paper's dimensionality of 1536 as its original embedding model has been deprecated. The paper's learning rate of 0.02 diverges here, so we tuned and selected $10^{-3}$ on validation performance. Following the original work we train with a binary cross-entropy objective, applied per outcome type.

In addition to these eight trained methods, we include a general-purpose frontier large language model, GPT-5.4~\cite{openai_gpt54}. The model is prompted directly with the names and SMILES strings of the two drugs and the candidate outcome, and constrained to answer only \texttt{Yes} or \texttt{No}. We then take the returned token log-probabilities (the probability the model assigns to answering \texttt{Yes}) and rank with it, so the same threshold-free AUROC, AUPRC, and Fmax are computed as for every other method. Because scoring requires one model call per (drug, drug, outcome) sample, exhaustive evaluation over the full test sets is not tractable; we therefore evaluate on an outcome-stratified sampled subset of each test split, drawing 5{,}000 positive samples with at least one sample per outcome and pairing each with the two negatives that the split itself prescribes. \name is re-evaluated on the subset as well, so the two models are compared on identical data. These numbers are not comparable to the full-test-set numbers reported for the other existing methods, and we report them separately in Supplementary Table~\ref{tab:gpt54_subset}. Full prompts, the scoring rule and the subset construction are given in Supplementary Note~\ref{note:gpt54}.

All existing methods are run from their public implementations, adapted as needed to our dataset splits, except LLM-DDI, for which no implementation is available and which we therefore reimplemented following the paper. For each existing method we start from the hyperparameters recommended by its authors or reference implementation and tune them if training does not converge on a certain dataset split; DDKG, MUFFIN, and TIGER use their published hyperparameters.

All methods are then evaluated under the protocol specified in Sections~\ref{method:benchmark_dataset_split} and \ref{method:negatives}; the same seeds as \name, namely five seeds (0, 1, 2, 42, 99) for split-by-drugs (ATC), split-by-drugs (targets), and split-by-drug pairs, and three seeds (0, 42, 99) across three independent repeats of the random split (nine runs in total); and the same macro-averaged AUROC, AUPRC, and Fmax. For every method, the reported test performance is computed from the checkpoint attaining the best validation AUPRC.
}

\section{Applications of \name}\label{method:case-studies}

DrugBank safety outcomes are annotated with one of nine organs, namely ``blood" (hematological), ``heart" (cardiovascular), ``liver" (hepatic), ``kidney" (renal), ``gastrointestinal", ``endocrine", ``urinary", ``immune", ``lung", or otherwise ``others/general" (Supplementary Table~\ref{tab:safety_outcome_organ}). Organs with fewer than five occurrences (``urinary", ``immune", ``lung") are not considered in all organ-level analyses. 7 out of 158 safety outcomes, including ``adverse effects, decrease", ``cardiotoxicity, decrease", ``hypertension, decrease", ``hypoglycemia, decrease", ``hypotension, decrease", ``nephrotoxicity, decrease", and ``therapeutic efficacy, increase" are considered as potentially beneficial safety outcomes and are thus excluded from all safety-oriented analyses. 
For each outcome $r$, we first rank the \name-predicted score $s_{(d_1, d_2, r)}$ among scores of all pairwise drug combinations between DrugBank small molecule compounds in our dataset for this outcome, then normalize it to [0,1]. Concretely, let $N$ be the number of drug pairs evaluated, let $\mathrm{rank}^{\downarrow}(\cdot)$ be the rank when sorting scores from largest to smallest among all $s_{(\cdot, \cdot, r)}$, then we obtain $1 - \frac{\mathrm{rank}^\downarrow(s_{(d_1, d_2, r)}) - 1}{N-1} \in [0, 1]$. We then derive \textit{prediction scores} by taking the geometric mean of these normalized ranks over five independently trained \name models and normalizing the result to [0, 1] in the same way. 
% \textit{prediction score} $\tilde{s}_{(d_1, d_2, r)}$

\subsection{Drug-Induced Effects on Liver, Heart and QT Prolongation}
\label{method:single_drug}
For each drug (drug A) in each dataset, we query the model trained on the DrugBank safety dataset with the input of the form (drug A, drug A, outcome) and obtain scores across all outcomes. For each outcome, we then rank the score among all scores of this outcome produced by 11,607 small molecule drugs or novel compounds (11,601 with DrugBank ID and 6 additional MASH clinical candidates) in our data using the same query format, before normalizing to [0,1]. \textit{Prediction scores} are the geometric mean of these normalized ranks over five independently trained \name models. 

When correlating our model predictions with annotations in each dataset, since the organ where the toxicity is measured differs, we also make sure the outcomes we consider for our model match such organs (Supplementary Table~\ref{tab:safety_outcome_organ}). 
Specifically, for the DILI (liver) dataset, we obtain predictions from our model with all liver-related outcomes (``excretion rate, increase $|$ serum level, decrease $|$ efficacy, decrease", ``liver damage, increase", ``liver enzyme elevations, increase", ``metabolism, decrease", and ``metabolism, increase"); for the DICT (cardiovascular) dataset, we obtain predictions from our model with all heart-related outcomes (in total, 44 outcomes); and for the DIQTA (QTc prolongation), we obtain predictions from our model with all QTc prolongation-related outcomes (``QTc prolongation, decrease", ``QTc prolongation, hypotension, increase", ``QTc prolongation, increase", ``QTc prolongation, torsade de pointes, cardiotoxicity, increase").
For DILI and DIQTA, we correlate annotations with predictions for each outcome individually; for DICT, due to the large number of outcomes, we correlate annotations with the average of the highest five predictions across 44 heart-related outcomes.

\subsection{Transporter, Carrier, and Enzyme-Mediated Outcomes}
\label{method:transportome}

We identify safety outcomes in the DrugBank dataset that are potentially transporter-mediated, including ``absorption, decrease", ``absorption, decrease $|$ serum level, decrease $|$ efficacy, decrease", ``absorption, increase $|$ serum level, increase $|$ adverse effects, increase", ``excretion rate, decrease $|$ serum level, increase", ``excretion rate, increase $|$ serum level, decrease $|$ efficacy, decrease", ``excretion, decrease", ``excretion, increase", ``serum level of the active metabolites, decrease", ``serum level of the active metabolites, decrease $|$ efficacy, decrease", ``serum level of the active metabolites, increase", ``serum level, decrease", ``serum level, increase". Among them, ``excretion rate, decrease $|$ serum level, increase", ``excretion, decrease", ``serum level of the active metabolites, increase", and ``serum level, increase" are considered as safety outcomes that are relevant to increase in serum concentration as explored in~\cite{Shi_2024_transportome}. 
We also identified safety outcomes in the DrugBank dataset that are potentially carrier-mediated, including ``protein binding, decrease", ``serum level, decrease", and ``serum level, increase". Similarly, potential enzyme-mediated safety outcomes include ``bioavailability, decrease", ``bioavailability, increase", ``metabolism, decrease", ``metabolism, increase", ``serum level of the active metabolites, decrease", ``serum level of the active metabolites, decrease $|$ efficacy, decrease", ``serum level of the active metabolites, increase".

To examine identified transporter-mediated DDIs validated in~\cite{Shi_2024_transportome}, we first query the model to obtain prediction scores for the above safety outcomes that are relevant to the increase in serum concentration between doxycycline and each of digoxin, warfarin, tacrolimus, and levetiracetam. Piracetam is a reference because it is structurally similar to levetiracetam with a side chain modification. It is known to interact with doxycycline, leading to a decrease in excretion and thus, increase in serum concentration. The maximum of the four prediction scores is presented for each drug pair. Since each drug can interact with many other substrates of their respective transporter (BCRP and MRP2 here), we also calculate two additional values: (1) the quantile of the maximum prediction score among all pairs of the form (doxycycline, X), and (2) the quantile of the maximum prediction score among all pairs of the form (digoxin, X), (warfarin, X), (tacrolimus, X), or (levetiracetam, X), calculated individually for each drug, where X is any other DrugBank compound we curate.

To systematically compare the maximum prediction score of transporter-mediated, carrier-mediated, and enzyme-mediated outcomes among drug pairs with and without overlap in their transporter, carrier, and enzyme profiles, respectively, we consider all drug pairs between drugs with respective profiles available in DrugBank and partition them into two groups, depending on whether or not the two drugs' profiles overlap. The maximum prediction score of each safety outcome group is then taken for each drug pair and aggregated according to the drug pair grouping. 

Finally, drugs that share each specific transporter are paired and queried to the model to probe into the potential outcomes mediated by individual transporters. The median across all such drug pairs is then taken to rank the relevance of each safety outcome. The signs of outcomes are neutralized. Representative carriers, enzymes, and targets with many drugs sharing them are taken as controls, with the outcomes ranked similarly.

\subsection{Drug Combination Clinical Trials}
\label{method:clinical_trials}
We derive pairwise safety comparisons from clinical trial AE data and compare them with those derived from \name predictions for the corresponding drug combinations. We restrict this analysis to AEs with $\geq$5 significant arm pairs across the curated trials, namely, alopecia, anemia, hypoglycemia, and neutropenia. 

Agreement between \name predictions and AE data for some adverse event $e$ is assessed by whether the safer arm in the trial also receives the lower \name score. More precisely,
\[
\mathrm{Agreed}_{e} \;=\;
\mathbbm{1}\!\Bigl\{
      \mathbbm{1}\bigl\{\mathrm{Incid}_{e}^{(1)} \ge \mathrm{Incid}_{e}^{(2)}\bigr\}
      \;=\;
      \mathbbm{1}\bigl\{\mathrm{Score}_{e}^{(1)} \ge \mathrm{Score}_{e}^{(2)}\bigr\}
\Bigr\}.
\]
where $\mathbbm{1}\{\cdot\}$ denotes indicator function, $\mathrm{Incid}_{e}^{(i)}$ denotes the incidence of adverse event $e$ for arm $i$, and $\mathrm{Score}_{e}^{(i)}$ denotes the \name predicted score for adverse event $e$. Because both DrugBank and TWOSIDES outcomes can be mapped to some of the AEs, we utilized models trained on both datasets and apply \name trained on DrugBank first to compare predicted scores. If the score difference is less than 0.1, we resort to \name trained on TWOSIDES for decision.

\subsection{Type 2 Diabetes Comorbidities}
\label{method:t2d_comorb}

We first query \name with pairs composed of pioglitazone or rosiglitazone and all T2D drugs of a different mechanism of action before taking the mean across all pairs of such drugs for each outcome related to myocardial infarction or stroke (Supplementary Table~\ref{tab:mi_stroke_outcomes}). 

For the hyperkalemia analysis, we consider ``hyperkalemia, increase", ``hypotension, hyperkalemia, nephrotoxicity, increase", ``renal failure, hyperkalemia, hypertension, increase", and ``renal failure, hypotension, hyperkalemia, increase" as hyperkalemia-related outcomes. We then query the model with pairs composed of each HF drug with all T2D drugs of a different mechanism of action. We take the maximum value across the above four outcomes as the safety score, representing each drug pair's level of hyperkalemia-specific safety concern. For each MoA group of HF drugs, all drug pairs containing an HF drug within the group are considered when plotting the point plot, using the geometric mean as the estimator. 

To generate a safety profile for each MASH drug or clinical candidate when combined with T2D drugs or clinical candidates, we first take the average of the highest five prediction scores for each pair (MASH drug or candidate, T2D drug or candidate). Then, for each MASH drug or clinical candidate, we take the lowest five drug pairs containing it with such scores. This effectively gives us a scalar score for each MASH drug or clinical candidate, representing the (worst) safety profile of the best possible combinations with T2D drugs containing it.

\subsection{MASH Combination Therapies}
\label{method:nash_combo}

For MASH drug combinations currently under clinical investigation, we first manually annotate them to be either ``efficacy" or ``safety" based on descriptions of the rationales of developing such drug combinations in existing literature~\cite{Suri_2022_nash_combos,Harrison_2023_nash_challenges,Tilg_2023_nash_challenges}. We then obtain the average of the highest five prediction scores. In addition, for each such combination, we take all drug pairs with aligned MoA pairs and calculate the average of the highest five prediction scores, treating those as reference safety for the combination.

\subsection{Drug Combination Prediction in BeatAML}
\label{method:beataml}

We use principal component analysis (PCA) to reduce the dimension of gene expression data from 22,783 to 150, which retains 90.3\% of variance.
We binary encode somatic mutation data, considering only pathogenic or potentially pathogenic mutations, and filter out those genes with fewer than three mutations across all patients. We then use multiple correspondence analysis (MCA) to reduce the dimension of the somatic mutation data (447 genes, 35 after filtering) to 30, which retains 92.9\% of inertia. 
We also keep clinical attributes with less than 10\% missingness and impute with either the most frequent or mean values, depending on whether the attribute is categorical or numeric. We exclude technical and administrative attributes and attributes about patient information after specimen collection. After filtering, the following attributes are kept: ``gender", ``ageAtDiagnosis", ``priorMalignancyNonMyeloid", ``cumulativeChemo", ``priorMalignancyRadiationTx", ``priorMDS", ``priorMDSMoreThanTwoMths", ``priorMDSMPN", ``priorMDSMPNMoreThanTwoMths", ``priorMPN", ``priorMPNMoreThanTwoMths", ``riskGroup", ``specificDxAtAcquisition", ``ageAtSpecimenAcquisition", ``specimenGroups", ``specimenType", ``FLT3\_ITDCall", ``NPM1Call", ``priorTreatmentTypeCount", ``priorTreatmentRegimenCount", ``priorTreatmentStageCount".

To adapt \name for personalized drug combination synergy prediction, we first use frozen \name encoders trained with the DrugBank combination safety dataset to generate drug embeddings. Then, we adopt a symmetric bilinear decoder to fuse the two drug embeddings. We concatenate the fused output with dimension-reduced gene expression, somatic mutation data, and clinical attributes before feeding into an MLP, which is trained from scratch to predict binary labels of whether or not drugs combinations are synergistic for the patient, defined above in Methods Sec.~\ref{method:beataml-dataset}. 
For each \name model (trained with five seeds), we use five seeds to train the bilinear decoder and the MLP for at most 200 epochs, leading to 25 models evaluated for each group. 
The model's hyperparameters are the bilinear decoder output dimension $= 128$, MLP hidden dimensions $= [256, 128]$, MLP dropout $= 0.2$, and learning rate $= 0.001$. The AdamW optimizer is used for training.

For evaluation, 10\% of patients or drugs are sampled and all associated response data are held out. We use AUROC as our primary performance metric. We calculate AUROC in two distinct ways: (1) Patient-centric AUROC: For each patient, we compute the AUROC across all drug combination predictions, then averaged these values over all patients. (2) Drug-centric AUROC: For each drug combination, we compute the AUROC across predictions from different patients, and then averaged these values.

\subsection{Drug Combination Response Prediction in Xenografts}
\label{method:pdx}

Given the small number of samples available, we use PCA to reduce the dimension of gene expression data from 20,684 down to 25, which retains 64.5\% of variance. \revision{A preprocessing ablation over six alternatives, including PCA to 60 and 100 dimensions, the top 25 and 100 highly variable genes, and the 25 and 100 highest-scoring genes by $F$-test, changes Kendall's $\tau$ by less than 0.05 in every case, so the reported PDX results are robust to this choice (Supplementary Note~\ref{note:pdx_robustness}; Supplementary Fig.~\ref{fig:supp_fig_7}e).}
We binary encode somatic mutation data, considering only pathogenic or potentially pathogenic mutations, and filtered out genes with fewer than three mutations in all patients. We then use MCA to reduce the dimension of the somatic mutation data (2,935 genes, 326 after filtering) to 25, which retains 63.7\% of inertia. 

To adapt \name for personalized drug combination response prediction, we first use frozen \name encoders trained with the DrugBank dataset to generate drug embeddings as with the BeatAML dataset. Then, given the small amount of data available, we reduce the \name drug embeddings to 25 principal components, adopt a simple element-wise max to fuse the two drug embeddings, and concatenate the fused output with dimension-reduced gene expression and somatic mutation data before feeding into a random forest regressor. 
For each \name weight (trained with five seeds), we again use five seeds to train the random forest regressor, leading to 25 models.
The hyperparameters of the model are: number of estimators (trees) $= 1000$, criterion to measure the quality of a split $= \texttt{friedman\_mse}$, maximum depth of the tree $= \texttt{unbounded}$, minimum number of samples required to split an internal node $= 2$.

Given the small number of unique drug combinations in the dataset, we evaluate model performance by leaving out one drug combination each time. The prediction cutoff of responsiveness is set to -20, aligning with~\cite{Gao_2015_pdx} (complete response (CR), partial response (PR), stable disease (SD), progressive disease (PD) according to the mRECIST criteria), and the threshold of minimum predicted responder or nonresponder is set to 5.

\subsection{Mortality, Readmission, and Adverse Events Prediction in a Longitudinal Event-Time Cohort}
\label{method:ehrshot_task}

\xhdr{Hospital re-admission} This task predicts whether a patient will be re-admitted to the hospital within 30 days after being discharged from a visit, using their historical visit records as input. 

\xhdr{All-cause mortality} This task predicts the mortality outcome of the visit following the patient's current visit (excluding the final visit), using the current visit as input. 

\xhdr{Adverse events} This task predicts the seriousness of each AE (thrombocytopenia, hyperkalemia, hypoglycemia, hyponatremia, anemia) during each visit, using all events prior to the AE timestamp during the visit as input. 

To evaluate performance, we partition patients for each task into train, validation, and test sets in an 80:10:10 ratio using a fixed random partition (random seed 42), without stratification. Because each patient contributes a single snapshot, the three sets are disjoint at the patient level, and the test set is never used for model selection. For the 30-day re-admission and all-cause mortality tasks, the train, validation, and test sets contain 3,442, 430, and 431 patients, respectively. For the AE prediction tasks, the train sets contain 3,571, 3,424, 3,503, 3,436, and 3,605 patients, the validation sets contain 446, 428, 438, 429, and 451 patients, and the test sets contain 447, 428, 438, 430, and 451 patients, for thrombocytopenia, hyperkalemia, hypoglycemia, hyponatremia, and anemia, respectively. We use AUROC as the evaluation metric. Each \name result shown in Fig.~\ref{fig:figure_7}h and Supplementary Table~\ref{tab:ehrshot_perf} is averaged over five \name models trained with different seeds.

\subsection{Adverse Events Prediction in an Oncology Cohort}
\label{method:dfci_task}

\xhdr{Population-level associations} We filter for regimens with $\geq$ 32 patients for sufficient statistical power. The patient number threshold comes from the exact binomial test~\cite{Hanley1983,Onakpoya2018}: with a true incidence of 5\%, a cohort of 32 patients gives about 80\% power to observe at least one event. This leaves us with combinations abiraterone + leuprolide, bicalutamide + leuprolide, capecitabine + temozolomide, carboplatin + etoposide, carboplatin + gemcitabine, carboplatin + paclitaxel, carboplatin + pemetrexed, cisplatin + etoposide, cisplatin + gemcitabine, cisplatin + pemetrexed, dabrafenib + trametinib, fluorouracil + leucovorin, fluorouracil + mitomycin, fulvestrant + palbociclib, gemcitabine + paclitaxel, letrozole + palbociclib. For every qualifying two-drug regimen we compute the observed incidence of each AE outcome and measure its correlation with the corresponding \name (\name trained on TWOSIDES) predicted score using Kendall's $\tau$. Within these retained regimens, we additionally adjust for age, gender, palliative intent, and race (by fitting multivariable logistic regression models, with age and the \name score standardized to zero mean and unit variance). Tumor tissue type is not used as a covariate here as regimens are typically uniquely indicated for a small set of tumors. The \name coefficients from these models represent the score's adjusted log-odds effect on AE risk. For the race feature, categories with fewer than 100 patients are also grouped. 

\xhdr{Patient-level predictions} We include all 3,577 patients in the cohort. We train random forest models to predict individual AE occurrence using \name drug embeddings combined with patient characteristics (age, gender, palliative intent, and race). 
Consistent with the population-level analysis, tumor tissue type is not used as a covariate. 
Performance is estimated with 5-fold stratified cross-validation. 
\name embeddings are reduced to 32 dimensions by exact PCA, retaining 99.9\% of variance. 
For every model, mean and std are computed across five random seeds, each being the average over five-fold cross-validation; for \name, each seed additionally averages over five \name models trained with different seeds.

\clearpage

\section*{References}

{
\customspacing{1}
\bibliographystyle{naturemag}
\bibliography{refs}
}

\clearpage
% Supplementary Information (compiled separately)
\includepdf[pages=-]{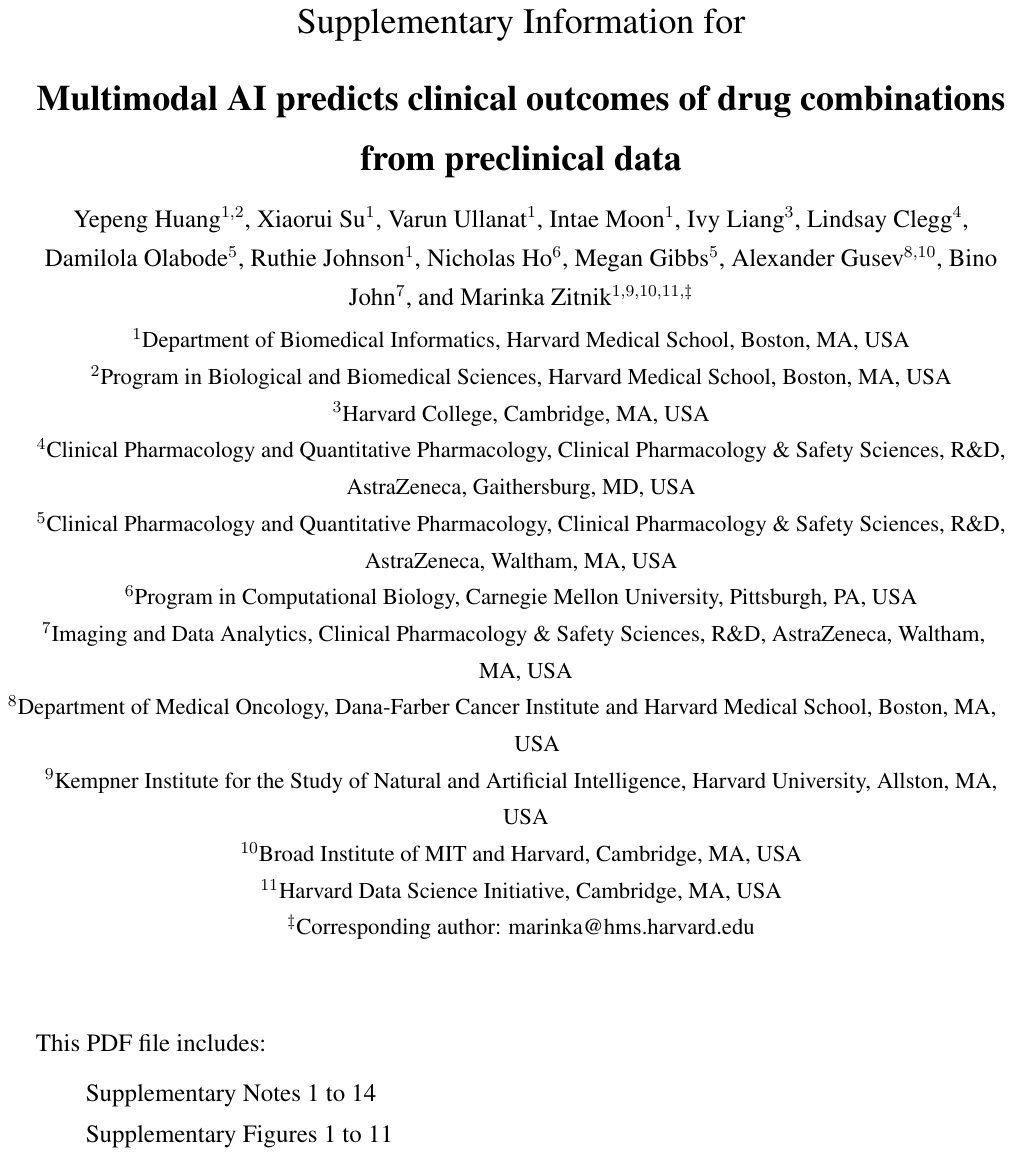}
\end{document}